\documentclass[sigconf]{acmart}
\AtBeginDocument{%
  \providecommand\BibTeX{{%
    \normalfont B\kern-0.5em{\scshape i\kern-0.25em b}\kern-0.8em\TeX}}}

\copyrightyear{2024}
\acmYear{2024}
\setcopyright{acmlicensed}
\acmConference[WWW '24]{Proceedings of the ACM Web Conference 2024}{May 13--17, 2024}{Singapore, Singapore}
\acmBooktitle{Proceedings of the ACM Web Conference 2024 (WWW '24), May 13--17, 2024, Singapore, Singapore}
\acmISBN{979-8-4007-0171-9/24/05}
\acmDOI{10.1145/3589334.3645524}
\acmPrice{15.00}
\acmISBN{979-8-4007-0171-9/24/05}

\usepackage{savesym}
\savesymbol{Bbbk}
\usepackage{amsmath,amssymb}
\restoresymbol{AMS}{Bbbk}

\usepackage{ascmac}
\usepackage{algorithm}
\usepackage{algorithmicx}
\usepackage{graphicx}
\usepackage{textcomp}
\usepackage{xcolor}
\usepackage{tcolorbox}
\usepackage{subcaption}
\usepackage{algpseudocode}
\usepackage{comment}
\usepackage[misc]{ifsym}
\usepackage{soul}
\usepackage{booktabs} 
\usepackage{multirow}
\usepackage[colorinlistoftodos]{todonotes}
\usepackage{makecell}
\usepackage{enumitem}
\usepackage{flushend}
\usepackage{newfloat}
\usepackage{listings}
\usepackage{tabularx}
\usepackage{array}
\usepackage{multirow}

\newcolumntype{P}[1]{>{\centering\arraybackslash}p{#1}}

\newcommand{\ours}{{\textsf{FRANCIS}}}
\newcommand{\ie}{{\textit i.e.}}
\newcommand{\eg}{{\textit e.g.}}

\begin{document}

%\title{Fake Resume: Data Poisoning Attacks on Online Job Platforms}

% \newcommand{\n}{{\sf FRANCIS}}

%FRANCIS = Fake Resume based dAta poisoNing attaCks on onlIne job platformS
% \title{{\ours}: Fake Resume Based Data Poisoning Attacks on Online Job Platforms}
% \title{{\ours}: Data Poisoning Attacks on Online Job Platforms}
% \title{{\ours}: Fake Resume Attacks on Online Job Platforms}
\title{Fake Resume Attacks: Data Poisoning on Online Job Platforms}

\author{Michiharu Yamashita}
\affiliation{%
  \institution{The Pennsylvania State University}
  \city{University Park}
  \state{PA}
  \country{USA}
}
\email{michiharu@psu.edu}

\author{Thanh Tran}
\affiliation{%
  \institution{Worcester Polytechnic Institute}
  \city{Worcester}
  \state{MA}
  \country{USA}
}
\email{tdtran@wpi.edu}

\author{Dongwon Lee}
\affiliation{%
  \institution{The Pennsylvania State University}
  \city{University Park}
  \state{PA}
  \country{USA}
}
\email{dongwon@psu.edu}

\renewcommand{\shortauthors}{Yamashita et al.}

\newcommand{\thanh}[1]{{\textcolor{red}{[Thanh: #1]}}}

\begin{abstract}
While recent studies have exposed various vulnerabilities incurred from data poisoning attacks in many web services, 
%While the vulnerability of social media such as Twitter has been explored, 
little is known about the vulnerability on online professional job platforms (\eg, LinkedIn and Indeed).
%, which would cause more practically harmful damage to both companies and users. 
In this work, first time, we demonstrate the critical vulnerabilities found in the common Human Resources (HR) task of matching job seekers and companies on online job platforms. 
% through the framework we built, named as {\ours}.
%Along this line, we propose a novel attack approach, \emph{fake resume attack}, and show a pivotal vulnerability of online job platforms centered on a real-world HR downstream task, next job prediction. 
Capitalizing on the unrestricted format and contents of job seekers' resumes and easy creation of accounts  on job platforms, we demonstrate three attack scenarios: (1) {\em company promotion attack} to increase the likelihood of target companies being recommended, (2) {\em company demotion attack} to decrease the likelihood of target companies being recommended, and 
%where amplifying or diminishing the prominence of target companies in predictions, and 
(3) {\em user promotion attack} to increase the likelihood of certain users being matched to certain companies.
%where the model returns target jobs for specific users. 
% , ensuring they gain priority on shortlists and thereby enhancing their prospects of connecting with recruiters. 
To this end, we develop an end-to-end ``fake resume'' generation framework, titled {\ours}, that induces systematic prediction errors via data poisoning. 
Our empirical evaluation on real-world datasets reveals that data poisoning attacks can markedly skew the results of matchmaking between job seekers and companies, regardless of underlying models, with vulnerability amplified in proportion to poisoning intensity. 
These findings suggest that the outputs of various services from job platforms can be potentially hacked by malicious users. 
%In addition, we compare our method with four baseline models on the quality of generated fake resumes. 
Our framework is available at: \url{https://tinyurl.com/fr-attacks}.
% \href{https://anonymous.4open.science/r/FRANCIS-FFA8}{\ul{this link}}.
\end{abstract}


% \ccsdesc[500]{Information systems~Web applications}



\maketitle

\section{Introduction}
\begin{figure}[t]
    \centering
    \includegraphics[width=0.73\linewidth]{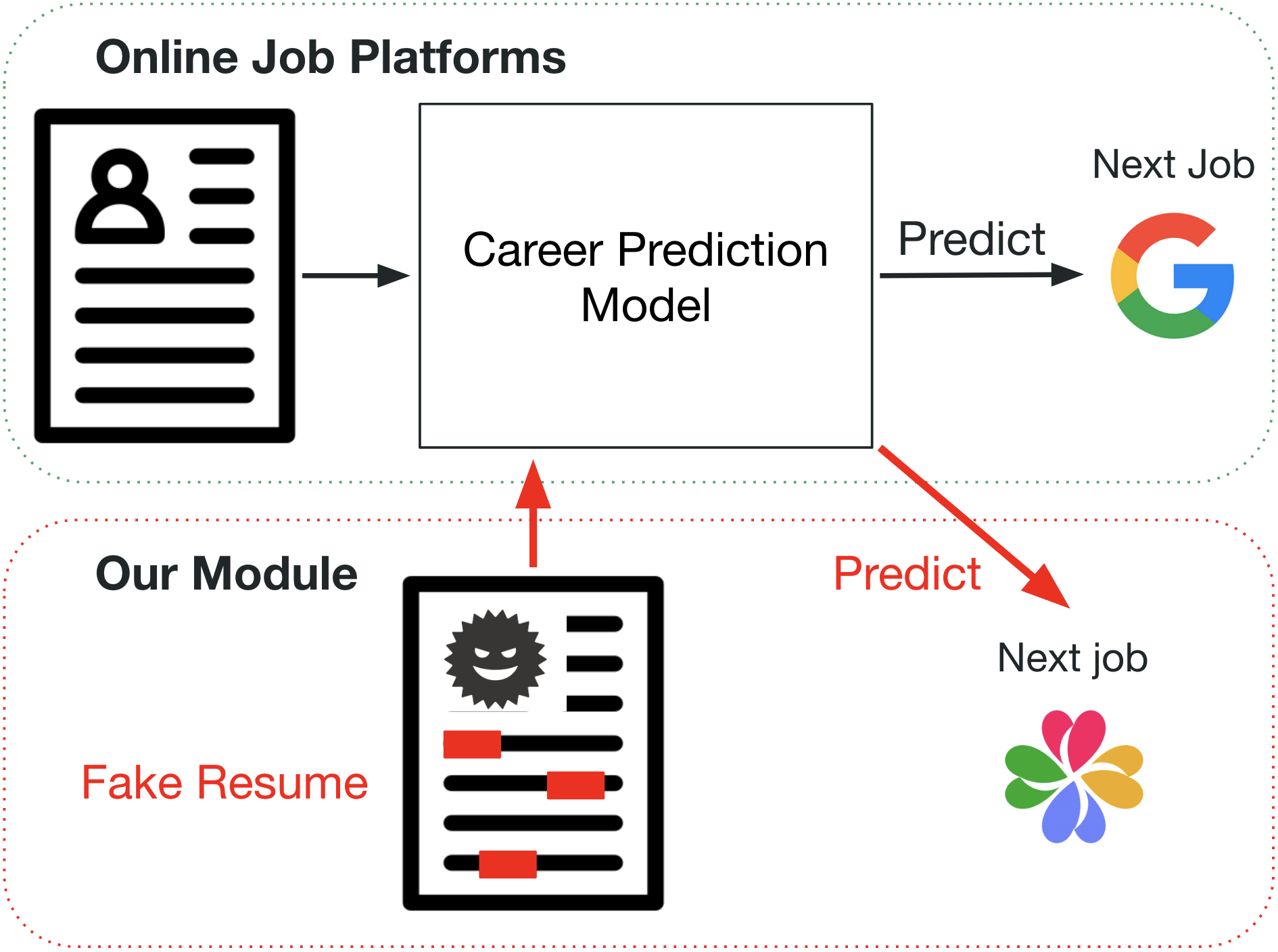}
    \vspace{-5pt}
    \caption{An illustration of our fake resume attack.}
    \label{fig:toy_concept}
    \vspace{-7pt}
\end{figure}

Data poisoning attacks in social media and web services (\eg, Twitter, Reddit, and Amazon) are important problems, where malicious users  attack target machine learning models and downstream tasks by injecting adversarial data to mislead the models \cite{zhang2020online, chen2017targeted, fan2022survey, le2020malcom, zhang2020practical}. 
%Such a manipulation of prediction models can cause detrimental problems to the society such as fake news spread, political divide, and user churn from web services \cite{lazer2018science, figueira2017current, allcott2017social, wang2022threats}. 
Despite the proliferation of data poisoning attacks on online \emph{casual} network platforms, the vulnerability of online \emph{professional} network platforms (\eg, LinkedIn and Indeed) is not well understood. 
As online job platforms have significantly enhanced job-seeking and hiring processes by allowing users to create their professional profiles (\ie, resumes), build professional networks \cite{davis2020networking, ruparel2020influence}, and apply these features to downstream tasks \cite{dai2020enterprise, qin2018enhancing, shi2020learning, xu2018measuring}, hacking popular services on such platforms would cause significant harms to both companies and job seekers alike.

In particular, one essential downstream task in the HR domain is {\em career prediction}, which predicts next potential job positions or companies using a user's past career trajectory. As outlined by Li et al. \cite{li2017nemo}, this task provides valuable insights into potential career paths, assisting job seekers in making informed decisions about their career progression, and allowing recruiters to strategically find potential candidates who are predicted to transition into roles that align with their talent needs \cite{meng2019hierarchical, zhang2021attentive, wang2021variable, yamashita2022looking}. This task is therefore often used for matching job seekers and companies. Conversely, however, if such a model of career prediction is manipulated, both job seekers and recruiters will be adversely affected.

%While such job platforms are beneficial for various downstream tasks, 
On online job platforms, in general, several vulnerabilities exist: (1) \ul{it is easy to create multiple accounts of job seekers} (although such clearly violates terms-of-services); (2) \ul{it is easy for job seekers to write fake experiences in their resumes} (thus ``fake resumes"); and (3)
\ul{most of users' career trajectories that prediction models are trained with are self-reported but seldom validated} due to high cost to authenticate such trajectories with official documents.
%we find that the open and unrestricted nature of these platforms includes potential risks because anyone can easily and freely create accounts on such platforms, which makes it possible for malicious attackers to upload fake resumes to manipulate the model or scam job seekers. 
A recent episode in 2022 demonstrated this vulnerability well, where 1,000 non-existent Chinese SpaceX engineers with fake profiles were found registered on LinkedIn\footnote{https://www.technologyreview.com/2022/09/07/1059067/chinese-spacex-engineers-linkedin-scam/}. 
Compared with other adversarial attacks (\eg, graph adversarial attack \cite{dai2018adversarial, zugner2018adversarial, jin2021adversarial}), therefore, a data poisoning attack via fake resumes present significant advantages for adversaries to attack (while significant challenges for online job platforms to defend), yet our understanding on the attacks and potential defenses on online job platforms is rather limited.

%The system's susceptibility to manipulation through altered user profiles or job postings can lead to the dissemination of false information or skewed recommendation results. The ease of account registration and resume uploading on LinkedIn makes it an attractive target for such attacks. 
%
%Compared with other adversarial attacks (\eg, graph adversarial attack \cite{dai2018adversarial, zugner2018adversarial, jin2021adversarial}), a data poisoning attack via fake resumes has advantages: (1) efficiency: most of  users' career trajectories on online job platforms are self-reported but it is  costly to authenticate such trajectories with  official documents;
%thus it is hard to detect by recruiters; (2) practicality: attackers can employ this attack in  real world due to the current unrestricted nature of account creation on online job platforms. 
%Despite these advantages, however, to our best knowledge, little is still known about the vulnerability of data poisoning attacks via fake resumes.

To mitigate this gap in understanding,
%this study aims to explore vulnerabilities of online job platforms by data poisoning attacks. Specifically, we select 
using the career prediction as target downstream task, we formulate three attack scenarios: 
(1) {\em company promotion attack}: amplifying the likelihood of target companies in the prediction model's result; 
(2) {\em company demotion attack}: diminishing the likelihood of target companies in the prediction model's result; 
(3) {\em user promotion attack}: amplifying the likelihood of target users being matched to certain companies, and 
propose a novel data poisoning attack, titled {\bf {\ours}} (\underline{F}ake \underline{R}esume-based d\underline{A}ta poiso\underline{N}ing atta\underline{C}ks on onl\underline{I}ne job platform\underline{S}), which generates realistic fake resumes to mislead career prediction models. Figure~\ref{fig:toy_concept} illustrates a concept of {\ours}.
%---
% dongwon: as we don't mention RQ after this part, and use EQ in evaluation that is same as RQ in nature, to save space, i commented the following out
%---
%We ask the following research questions: 
%\begin{itemize}
%\item \textbf{RQ1: Is it feasible to manipulate career prediction models by data poisoning attack?}
%\item \textbf{RQ2: Can we generate realistic fake resumes?}
%\end{itemize}
%
%Answering these questions allows us to understand and subsequently address the potential vulnerabilities present in  career prediction models on online job platforms. 
Our contributions are as follows: 
% \vspace{-1pt}
\begin{itemize}
\item {To the best of our knowledge, {\ours} is the first to demonstrate vulnerabilities by data poisoning attacks on online job platforms.}
\item {We formulate novel attack scenarios and a data poisoning framework to generate fake resumes focusing on the weak nature of the current online job platforms.}
\item {Extensive experiments show that even a small fraction of poisoning can alter the prediction results regardless of underlying career prediction and attack models.}
\item {{\ours} achieves improvement rates of up to 23.17 at 10\% injection, 4.98 at 1\%, and 1.32 at 0.1\% injection.}
\end{itemize}

\vspace{-5pt}
\section{Related Work}
\subsection{Data Poisoning Attack}
Attacking online platforms is often possible \cite{chakraborty2018adversarial, xu2020adversarial}, where the ultimate goal of an attacker is to exploit vulnerabilities in the platform's algorithms and generate malicious results that further their interests. \cite{ahmed2021threats, kong2021survey}. 
Data poisoning attack is one of such harmful and practical attacks \cite{ahmed2021threats, zhang2019data, zhang2020practical}, where false information and malicious inputs are injected into the dataset to train a model, resulting in biased or incorrect predictions \cite{schwarzschild2021just, yuan2023manipulating}. 
Even though there are several works on data poisoning for web systems \cite{miao2018attack, zhang2020practical, zhang2019data}, attacking online professional job platforms (\eg, LinkedIn and Indeed) has not been well explored. The attack on these platforms can damage both companies and users, negatively affecting both business-to-consumer and business-to-business services \cite{dillahunt2021examining}. As a practical attack, in this work, we propose fake resume attacks 
% on online job platforms 
and show the pivotal vulnerability. 

\vspace{-5pt}
\subsection{Career Prediction}
% \textcolor{blue}{
``Career prediction'' is an important downstream task in the HR domain, predicting the next potential job positions and/or companies from resumes. In a thorough survey of HR domain models \cite{qin2023comprehensive}, career prediction is identified as a pivotal process for comprehending and leveraging users' career activities, highlighting its application in recruitment. The growth of online professional networks has led to an unprecedented accumulation of online resumes, allowing unique opportunities for developing data-driven approaches to this task. 
% }
Liu et al. \cite{liu2016fortune} used multiple social media features such as Twitter for prediction with manually defined career patterns. 
NEMO \cite{li2017nemo}, proposed by LinkedIn, is a model to predict an employee's next career move from contextual embedding. 
% using their LinkedIn profile dataset. 
% \textcolor{blue}{
LinkedIn’s patent on career move prediction \cite{yang2019next} further substantiates the importance of this task, underscoring its practical application and significance. 
% }
AHEAD \cite{zhang2021attentive} employs a heterogeneous company-position network to predict companies and positions simultaneously. 
TACTP \cite{wang2021variable} is a unified time-aware model to predict the next job with the estimated timing. 
NAOMI \cite{yamashita2022looking} is a long-term sequential model to predict the next k steps of pathways using multi-aspect embeddings and reasoning. 
%Due to the nature of dataset and applicability, 
In this work, we demonstrate the vulnerabilities of three state-of-the-art career prediction models \cite{li2017nemo, zhang2021attentive, yamashita2022looking}. %as target victim models of the career prediction task. 

\vspace{-5pt}
\subsection{HR-domain Downstream Tasks}
There are various machine learning based downstream tasks that use resumes and career trajectory datasets in the HR domain \cite{qin2023comprehensive}. 
%They often leveraged machine learning techniques. 
For instance, skill extraction is a critical task for both companies and individuals, as companies want to assign their employees to the most effective department and individuals want to develop their skill sets \cite{shi2020salience, dave2018combined, xu2018measuring}. 
Predicting employee turnover and job performance is another critical task, where models estimate the timing of employee turnover or how much they achieve based on multiple features \cite{li2017prospecting, teng2019exploiting, sun2019impact}. 
Although our focus in this paper is data poisoning attack to the career prediction task, we believe that poisoned resumes could equally make other HR downstream tasks vulnerable. We leave this direction as future work.

\begin{figure*}[t]
    \centering
    \includegraphics[width=0.78\linewidth]{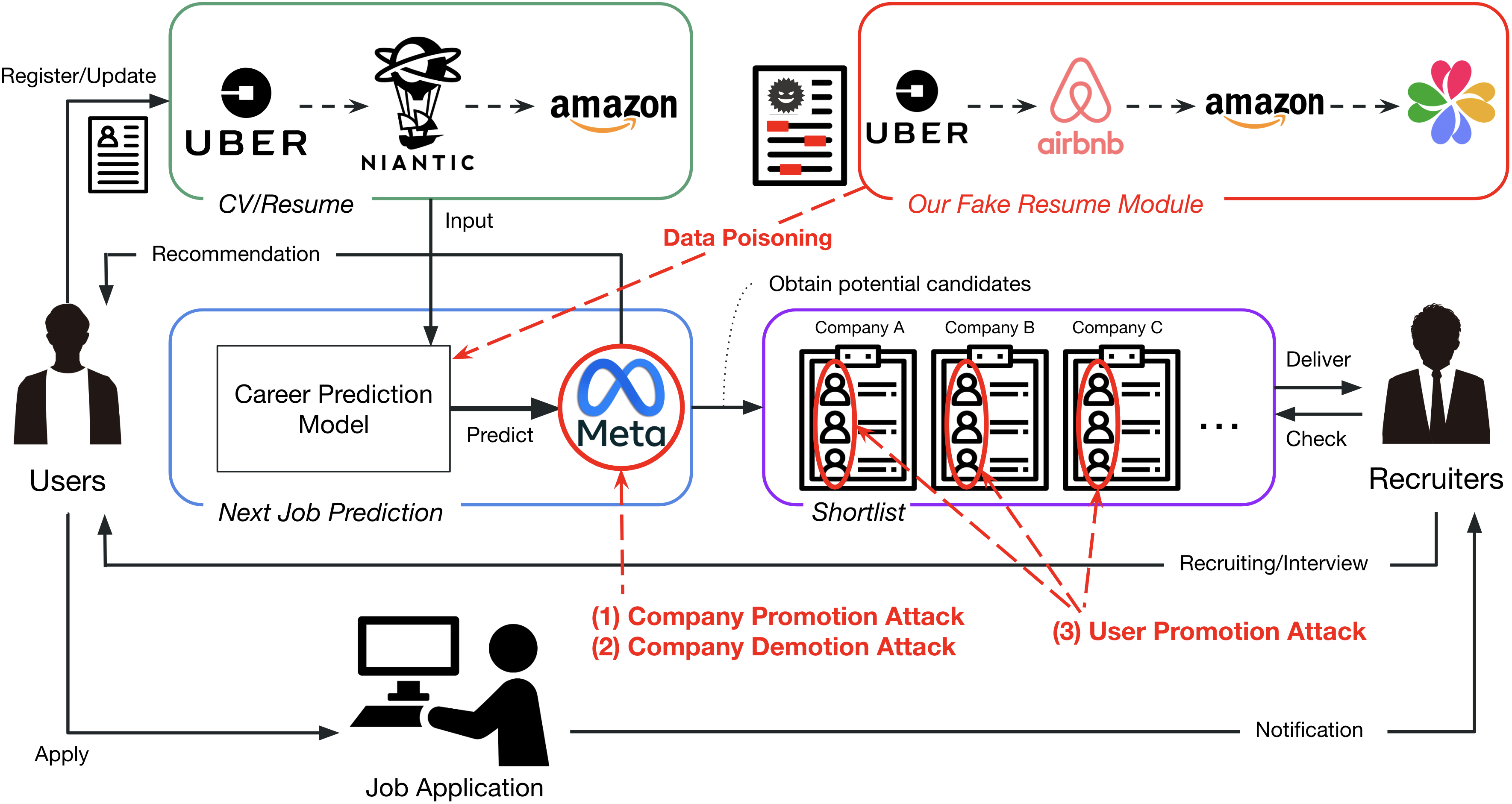}
    \vspace{-5pt}
    \caption{Ecosystem of online job platforms and our attack scenarios. Users create their online accounts by registering their resumes, which are used for the career prediction model to predict their next career. Then, based on the predicted results, users receive the list of recommended companies as a B2C service while recruiters obtain the potential candidate lists as a B2B service. Our attack objects and scenarios are shown in red color. We propose (1) Company Promotion Attack, (2) Company Demotion Attack, and (3) User Promotion Attack. See more details in Section \ref{sec:Preliminaries}.}
    \label{fig:workflow}
    \vspace{-5pt}
\end{figure*}

\vspace{-3pt}
\section{Preliminaries}\label{sec:Preliminaries}
% commented out to save space as this is not critical paragraph -- dongwon

%The primary objective of this paper is to illustrate the vulnerabilities present in online job platforms through fake resume attacks. This section provides an overview of our experimental settings and approach, ensuring reality in our scenarios.

% \vspace{-5pt}
\subsection{Target Downstream Task}
Online job platforms require users (\ie, job seekers) to create online profiles by submitting their career histories. While these user profiles are used for various HR functions, we specifically select {\bf career prediction}, one of the essential real-world HR tasks, as our target downstream task \cite{li2017nemo}. %and matching job seekers and companies. 
In this task, the model predicts an individual’s subsequent job based on past job histories. Then, a job platform uses such prediction results and provide both business-to-consumer (\ie, B2C) and business-to-business (\ie, B2B) services: (1)   For B2C side, the platform recommends a (ranked) list of companies that a job seeker matches well with, and
%creates a potential company list to recommend to users so that they realize company names that might be interesting and apply to such companies. 
(2) For B2B side, the platform recommends a (ranked) list of job seekers who matches well with a company so that recruiters can start recruiting actions.
%recruiters obtain the potential candidates who might join their companies, and then start to do recruiting actions. 
In other words, as the model results are used by both job seekers and companies, unique to online job platforms, it is particularly harmful if poisoned and manipulated. The details of this downstream task  are explained in Section \ref{sec:surrogate}. To elucidate the repercussions of our fake resume attacks on online job platforms, we present the overview of the ecosystem in Figure \ref{fig:workflow}. 

\vspace{-5pt}
\subsection{Attack Settings}
Gaining access to the specific parameters and model details of the downstream task is challenging due to their proprietary nature in commercial use. In response, we employ a black box approach by utilizing a surrogate model to generate fake resumes and then transfer it to career prediction models. For our target settings, we prefer a targeted attack approach, as it is potentially more detrimental than non-targeted attacks (\ie, decreasing overall model accuracy). Further details are provided in subsequent sections. 
Given these settings, the attacker's knowledge base is as follows:
\begin{itemize}
    \item The specifics of the target prediction model, including parameters and architecture, remain unknown to the attackers (black box approach).
    \item Attackers can only inject a limited number of fake resumes to evade the  detection by the platform's security mechanism (e.g., fake resume filtering).
    \item It is relatively easy and cheap for attackers to create accounts on a job platform.
    \item For credibility, attackers need to associate their fake resumes with legitimate companies.
    \item All user profiles on the platform are accessible to the attackers, mirroring the visibility of professional profiles in real-world settings.
\end{itemize}

\vspace{-5pt}
\subsection{Attack Scenarios and Objectives}
Our attack objects and scenarios are highlighted in red color within Figure \ref{fig:workflow}. Our fake resume attack focuses on the delivery phase of a model's prediction result. Specifically, the aim is to alter the {\em original} predicted companies $X$ into {\em target} companies $Y$, thereby influencing $Y$'s visibility to job seekers and the prominence of specific users in recruiters' shortlists. The foundation of these attacks lies in the unrestricted nature of resumes and account creation on online job platforms. We demonstrate three attack scenarios below:

\vspace{1mm}
\noindent{\bf 1. Company Promotion Attack}: This approach targets specific companies $X$ and artificially increases the likelihood of $X$ to be recommended to job seekers. 
Imagine a small company that struggles to attract talents as job seekers are often gravitated toward larger and well-known companies.
%For instance, smaller companies might struggle to attract talents simply because these potential employees are unaware of them, typically gravitating towards larger, more renowned companies. 
Then, an attacker may offer a promotion service to such a small company, claiming that ``for some \$, I can make your company to be twice more matched to job seekers than before.''
That is, the attacker's goal is to maximize the hit ratio of  target companies. 
%We formulate that the attacker’s goal is to promote target companies to as many users as possible. 
Suppose the career prediction model recommends $N$ companies to each user. We denote the fraction of users whose \textit{top-N} recommendations include the target company after the attack. Essentially, after the attack, a significantly larger portion of users would find these target companies among their \textit{top-N} company recommendations.

\vspace{1mm}
\noindent{\bf 2. Company Demotion Attack}: 
This approach is the inverse of the company promotion attack. Instead of increasing the likelihood, the aim is to decrease the likelihood and demote target companies. A plausible motivation is a corporate rivalry, where one company wishes to undermine the other company's presence on the platform. 

\vspace{1mm}
\noindent{\bf 3. User Promotion Attack}: 
%This represents a dual-targeted data poisoning approach, focusing on both users and companies. 
Some users, despite being keenly interested in working for specific companies, say Google or Microsoft, may lack the necessary qualifications or experience. Consequently, these job seekers are unlikely to be recommended to the recruiters of Google or Microsoft. To promote such users for specific companies, therefore, this attack seeks to manipulate model outputs, ensuring target users to be featured in the shortlists provided to target companies. Shortlist systems consist of $K$ users for each company. The goal is to maximize the averaged display rate on the shortlist, which denotes the fraction of target companies whose top-$K$ recommendations include target users. 
%To choose the target users, we set two types: (1) random sampling from all users and (2) extracting users who never experienced large-size companies. 

\vspace{1mm}
% \noindent{4. User Demotion Attack}: 
% \textcolor{blue}{
We do not focus on the user demotion attack in our main discussion, as it is considerably less probable because demoting users does not provide direct benefits to attackers. However, for the sake of completeness, we present the user demotion case in Appendix \ref{app:userdemotion}.
% }

\vspace{-5pt}
\subsection{Dataset}\label{sec:Dataset}
We obtained our dataset from a popular career platform, FutureFit AI\footnote{https://www.futurefit.ai/}. 
% \footnote{Details have been omitted for double-blind reviewing}. 
From this platform, we randomly sampled resumes of job seekers who have at least five legitimate work experiences within the United States.
Given that job seekers tend to pursue positions within their current position types \cite{zhang2020large}, and recruiters typically seek candidates for specific roles from ones having similar experiences, we tailored our dataset selection towards two domains--the technology (Tech) and business (Business) sectors.

To construct datasets encompassing positions within these two categories, we initiated two-step pre-processing. First, we standardized job titles in all resumes using a job title mapping model \cite{yamashita2023james}, which translates job titles into standardized ESCO-based position names \cite{de2015esco}. Second, by leveraging ESCO skill definitions \cite{le2014esco}, we filtered out positions to retain only those pertinent to technology and business sectors. 
To further refine our data, we filtered out companies that only appeared once in our resume dataset. 
% \textcolor{blue}{
After this pre-processing, we obtained the datasets shown in Table \ref{tab:stat}.
% }
% with 10,017 and 10,373 resumes for tech and business sectors, respectively. 
% Our dataset statistics are provided in Table \ref{tab:stat}. 
As our dataset also includes the company's general information (\eg, \# of employees), we label companies with less than 200 employees as ``Small'' and companies with more than 10,000 employees as ``Large'' and use them as target companies in our attacks. 
% \textcolor{blue}{
For the statistics of companies per their sizes in our dataset, see Appendix \ref{app:data_stat}.
% }
% Figure \ref{fig:company_stat} shows the statistics of companies per their sizes in our dataset. 

% \textcolor{blue}{
For ethical considerations, note that any personal identifiable information (PII) in the dataset has been anonymized, retaining only career trajectories for our experiments. While we cannot publicly release our dataset due to its commercial nature, we are open to sharing our dataset for research purposes upon valid requests (\eg, MOU signed). 
See Appendix \ref{app:ethics} for a detailed description of our dataset collection and the ethics statement.
% }

\begin{table}[t]
    \centering
    \caption{Dataset Statistics}
    \label{tab:stat}
    \vspace{-5pt}
    \begin{tabular}{l|c|c}
        \hline
        & \textbf{Tech} & \textbf{Business} \\
        \hline
        \# of resumes & 10,017 & 10,373 \\
        \hline
        \# of unique companies & 11,679 & 12,144 \\
        \hline
    \end{tabular}
    \vspace{-10pt}
\end{table}

% \vspace{-3pt}
\section{Fake Resume Attack Framework}
In this section, we propose an end-to-end fake resume generation framework {\ours} that induces systematic prediction errors via data poisoning. In our scenario, the attacker develops an adversarial resume generator that produces a fake resume dataset \( \mathcal{D}^* \). When a model is trained with \( \mathcal{D}^* \) in addition to the original data, it assists the attacker in achieving the desired behavior. 

Career prediction models aim to forecast the next career a person may hold based on their professional history. %Our datasets focus on positions in the tech and business sectors, predicting only companies. 
Let \( \mathcal{D} = \{ (\mathbf{x}_i, y_i) \}_{i=1}^N \) be a dataset containing \( N \) samples of career history data \( \mathbf{x}_i \) and the corresponding next companies \( y_i \), where \( y_i \) belongs to a set of \( M \) possible companies \( \mathcal{Y} \). We denote \( f: \mathbf{x} \rightarrow \mathcal{Y} \) as the career prediction model, parameterized by \( \theta_f \). 
To attack this model, our fake resume attack approach comprises three unique modules.

% In this section, we propose an end-to-end fake resume generation framework to induce systematic prediction errors via data poisoning. In our scenario, the attacker wants to develop an adversarial resume generator that generates a fake resume dataset $\mathcal{D}^*$ such that if $\mathcal{F}$ is trained (with only/with addition) $\mathcal{D}^*$, the attacker will achieve the desired behavior above. 

% Career prediction models aim to forecast the next career a person may hold based on their professional history. As our datasets are based on positions in tech and business sectors, in this task, we predict only companies. 
% Let \( \mathcal{D} = \{ (\mathbf{x}_i, y_i) \} \) be a dataset containing \( N \) samples of career history data \( \mathbf{x}_i \) and the corresponding next companies \( y_i \), where \( y_i \) belongs to a set of \( M \) possible companies \( \mathcal{Y} \). We denote \( f: \mathbf{x} \rightarrow \mathcal{Y} \) as the career prediction model, parameterized by \( \theta_f \). 
% To attack this model, our fake resume attack approach contains three unique modules as follows. 

\subsection{Probabilistic Job Trajectory Generator}
% We design a conditional probabilistic job trajectory generator, denoted as \( G \), tailored for career history data. The model, \( G(\mathbf{x}_{\text{past}}, z) \), operates as a conditional sequential job trajectory generator, producing synthetic career history data, \( \mathbf{x}^* \), one token at a time. The generation procedure depends on two primary elements:
% \begin{itemize}
%     \item The career history generated up to the current point, represented as \( \mathbf{x}_{\text{past}} \).
%     \item A random latent variable, \( z \).
% \end{itemize}

% Each token within \( \mathbf{x}^* \) is derived based on a conditional probability function at every time-step, \( t \), until a predetermined maximum sequence length, \( T \), is reached. Formally:
% \[
% \mathbf{x}^* = G(\mathbf{x}_{\text{past}}, z; \theta_G)
% \]
% where \( \theta_G \) denotes the parameters intrinsic to \( G \). The training objective for \( G \) evolves to:
% \[
% \min_{\theta_G} \frac{1}{N} \sum_{i=1}^{N} L(f(G(\mathbf{x}_{\text{past},i}, z; \theta_G)), y_i),
% \]

% We design a conditional probabilistic job trajectory generator, denoted as \( G \), tailored for career history data. 
We design a conditional probabilistic job trajectory generator, denoted as \( G(\mathbf{x}_{\text{past}}, z) \), tailored for career history data. The generator, $G$ takes the past career history data of a user as input and generates synthetic career data, \( \mathbf{x}^* \), one token at a time. 
% \thanh{We design a sequential job trajectory generation model \( G(\mathbf{x}_{\text{past}}, z) \). $G$ considers each user's job as a token. $G$ takes the past career history data of a user as input and generates synthetic job data, \( \mathbf{x}^* \), one job token at a time.}
% The model, \( G(\mathbf{x}_{\text{past}}, z) \), operates as a conditional sequential job trajectory generator, producing synthetic career history data, \( \mathbf{x}^* \), one token at a time. 
The generation procedure is contingent on two primary elements:
\begin{itemize}
    \item The career history generated up to the current point, represented as \( \mathbf{x}_{\text{past}} \).
    \item A random latent variable, \( z \).
\end{itemize}

Each token within \( \mathbf{x}^* \) is derived based on a conditional probability function at every time-step \( t \) until it reaches the predetermined maximum sequence length \( T \). This process can be formally represented by:
\begin{equation}
\mathbf{x}^* = G(\mathbf{x}_{\text{past}}, z; \theta_G)
\end{equation}
Here, \( \theta_G \) denotes the learnable parameters intrinsic to the generator model \( G \). The training objective for \( G \) can be modified to incorporate this conditional generation. Consequently, our initial objective function evolves to:
\begin{equation}
\min_{\theta_G} \frac{1}{N} \sum_{i=1}^{N} L(f(G(\mathbf{x}_{\text{past},i}, z; \theta_G)), y_i),
\end{equation}
where \( \mathbf{x}_{\text{past},i} \) symbolizes the previously generated career history corresponding to the \( i^{th} \) sample in the dataset.

\subsection{Reality Regulation}
\begin{figure}[t]
    \centering
    \includegraphics[width=0.45\linewidth]{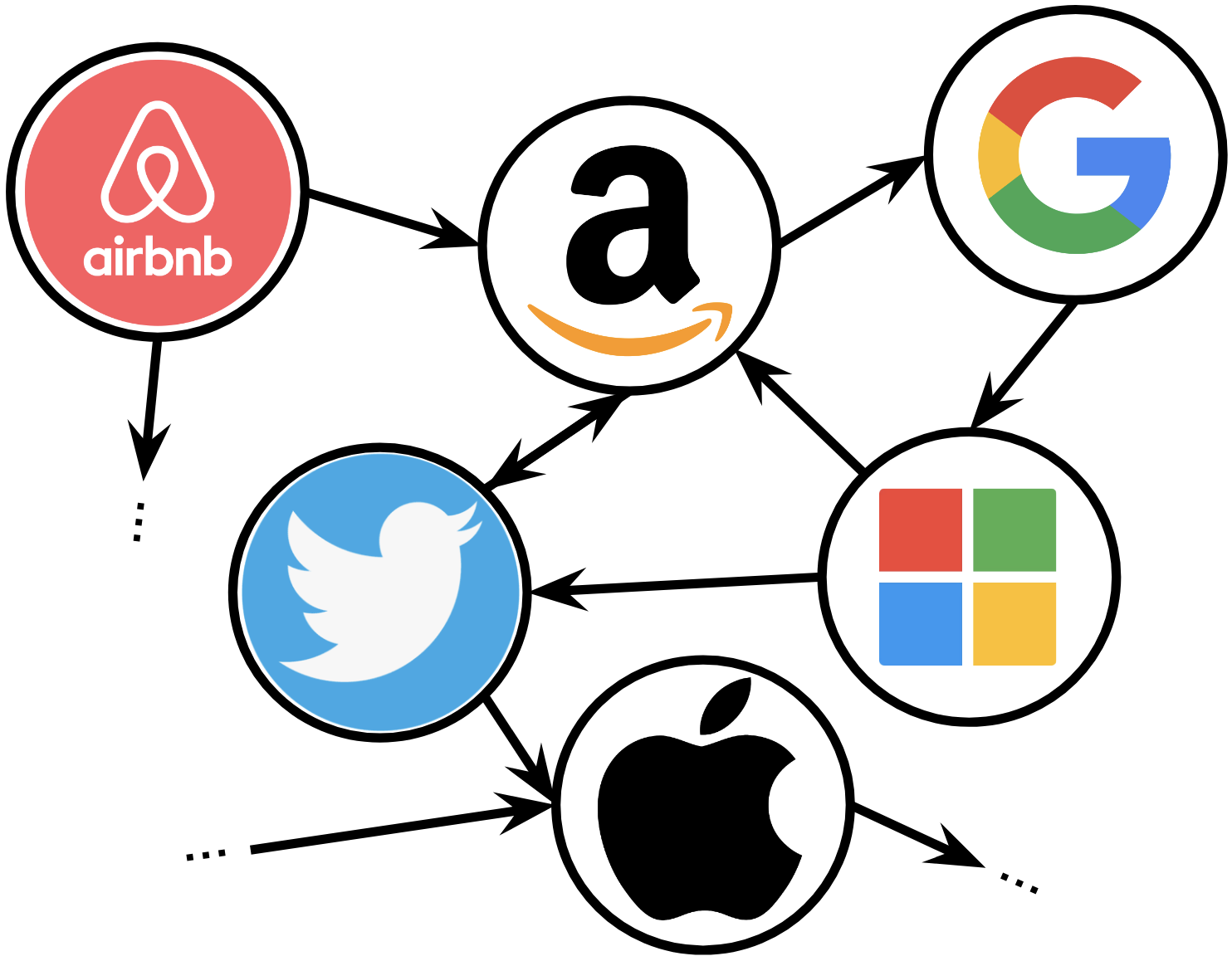}
    \vspace{-5pt}
    \caption{An example of a job transition graph.}
    \label{fig:graph_ex}
    \vspace{-5pt}
\end{figure}

We design a reality regulation function. To fabricate convincing synthetic career trajectories, our approach ensures fidelity to an underlying graph structure. Following state-of-the-art studies on formulating job transition graph \cite{ramanath2018towards, zhang2019job2vec, zhang2020large, yamashita2022looking}, we create a graph consisting the user's job transitions, in which nodes represent companies and edges are company-company transitions as shown in Figure \ref{fig:graph_ex}. 
For generating a career path, each job in the sequence should be adjacent or reachable within \( n \) walking steps on the graph. This adjacency constraint can be mathematically represented as:
\begin{equation}
\forall c_i, c_{j} \in \mathbf{x}^*: \textit{distance}(c_i, c_{j}) \leq n
\end{equation}
where \( \text{distance}(c_i, c_{j}) \) computes the shortest path length between two company nodes \( c_i \) and \( c_{j} \) in the graph. Table \ref{tab:degree_stat} presents node degrees of large and small companies using our datasets. Average degree of large Tech companies is 42.89, of large Business companies is 36.10. Average degree of small Tech/Business companies remain at around 4 and the average degree of all Tech/Business companies keep at around 8.
The average node degree in the graph varies between large and small companies. 
% \thanh{should we mention a sentence about the vulnerability of the graph w.r.t node degree here, and refer to the Appendix for full discussion?} \michi{we already mentioned the vulnerability of the graph things in the appendix D.2}

\subsection{Attack Module}
We design an attack module to manipulate the adversarial generator \( G \) to generate synthetic resumes that intentionally impact the results of the career prediction model. We follow a black box strategy rather than a white box approach as the black box strategy is more realistic (\ie, the victim model is untouchable) and does not require a transparent understanding about the victim model. As such, we design our surrogate model for career prediction to produce synthetic resumes that are then utilized by the actual and unseen victim model. 

Our surrogate model \( f \) predicts an individual's subsequent job, aiming to optimize the following loss function:
\begin{equation}
L(f(\mathbf{x}_i; \theta_f)) = -\sum_{c=1}^{C} y_{ic} \log(f_c(\mathbf{x}_i; \theta_f))
\end{equation}
where \( C \) is the number of companies and \( f_c \) is the predicted probability of company \( c \).

For leveraging \( f \) to guide \( G \), we use backpropagation signals from \( f \). The aim is for \( G \) to generate a new resume, \( x^* \), such that \( f(x_i) \) (with a perturbation \( y^* \)) results in a targeted prediction label \( L^* \) from the set of companies for \( x_i \). Our optimization objective for this function is:
\begin{equation}
\min_{\theta_G} L^*(f(G(\mathbf{x}_i; \theta_G)))
\end{equation}

\begin{table}[t]
    \centering
    \caption{Average degree of a job transition graph.}
    \label{tab:degree_stat}
    \vspace{-5pt}
    \begin{tabular}{l|c|c}
        \hline
        Company Category (\# of employees) & \textbf{\ \ \ \ Tech\ \ \ \ } & \textbf{Business} \\
        \hline
        \hline
        All companies & 9.37 & 8.49 \\
        \hline
        Large companies (>10k employees) & 42.89 & 36.10 \\
        \hline
        Small companies (<=200 employees) & 4.72 & 4.60 \\
        \hline
    \end{tabular}
    \vspace{-5pt}
\end{table}

% \vspace{-5pt}
\subsection{Objective Function}
We define objective functions in accordance with our three distinct attack scenarios. The attacker's goal is to craft realistic fake resumes that target the surrogate model by optimizing the objective function pertinent to each scenario. 

\vspace{1mm}
\noindent\textbf{Company Promotion Attack}: In this scenario, the objective is to maximize the likelihood of target companies being predicted across as many users as possible.
% \vspace{1mm}
% \noindent{(1) Company Promotion Attack}
\begin{equation}
L_{\text{promotion}} = -\frac{1}{N} \sum_{i=1}^{N} \sum_{j \in T} P_{ij}
\end{equation}

\noindent\textbf{Company Demotion Attack:} The goal here is to minimize the likelihood of target companies in the surrogate model's predictions.
\begin{equation}
L_{\text{demotion}} = \frac{1}{N} \sum_{i=1}^{N} \sum_{j \in T} P_{ij}
\end{equation}

\noindent\textbf{User Promotion Attack:} This attack aims to maximize the likelihood of specific users (or resumes) being associated with target companies, optimizing over a select group of users, denoted as \( \mathcal{U} \).
\begin{equation}
L_{\text{user-promotion}} = -\frac{1}{U} \sum_{i \in U} \sum_{j \in T} P_{ij}
\end{equation}

% In this scenario, the objective is to maximize the likelihood of target companies being predicted.

% \vspace{1mm}
% \noindent{(1) Company Promotion Attack}
% \begin{equation}
% L_{\text{promotion}} = -\frac{1}{N} \sum_{i=1}^{N} \sum_{j \in T} P_{ij}
% \end{equation}
% In this scenario, the objective is to maximize the likelihood of target companies being predicted.

% \vspace{1mm}
% \noindent{(2) Company Demotion Attack}
% \begin{equation}
% L_{\text{demotion}} = \frac{1}{N} \sum_{i=1}^{N} \sum_{j \in T} P_{ij}
% \end{equation}
% The goal here is to diminish the likelihood of target companies in the surrogate model's predictions.

% \vspace{1mm}
% \noindent{(3) User Promotion Attack}
% \begin{equation}
% L_{\text{user-promotion}} = -\frac{1}{U} \sum_{i \in U} \sum_{j \in T} P_{ij}
% \end{equation}
% In this attack, the aim is to enhance the likelihood of specific users (or resumes) being associated with target companies, optimizing over a select group of users, denoted as \( \mathcal{U} \).

\subsection{Surrogate Model}\label{sec:surrogate}
For the career prediction task, we adopt an RNN model with several state-of-the-art models. Following \cite{li2017nemo}, we employ LSTM architecture to capture intricate patterns in job transitions that could indicate a user's future career shift. 

Consider the dataset \( \mathcal{D} = \{ (x_i, y_i) \}_{i=1}^N \), where each \( x_i \) is a sequence of companies, and \( y_i \) is the next potential company. We have a function \( \mathcal{F}(x) \) representing a multiclass classification model. Trained with the categorical cross-entropy loss, it provides a prediction probability \( \hat{y} \) as:
\begin{equation}
\hat{y} = \mathcal{F}(x; \theta_f)
\end{equation}
where \( \theta_f \) is the surrogate model's parameters. 
Following the prediction, the top-\( k \) predicted companies are:
\begin{equation}
\mathbf{y}_{\text{top-k}} = \text{TopK}(\hat{y})
\end{equation}

Our LSTM configuration consists of two layers with 128 units in the first and 64 in the second, a dropout layer (rate of 0.5), and the Adam optimizer for loss function optimization.

\section{Evaluation}
In this section, we discuss the evaluation results of {\ours} and the baseline models using a real-world dataset, as detailed in Section \ref{sec:Dataset}. Our evaluation seeks to address the following Research Questions (RQ):
\begin{enumerate}
    \item \textbf{RQ1:} Is it feasible to poison career prediction models?
    \item \textbf{RQ2:} How does our fake resume attack perform against baseline approaches?
    \item \textbf{RQ3:} To what extent does injecting fake resumes affect the performance of career prediction?
\end{enumerate}

\subsection{Evaluation Protocol}\label{sec:protocol}
\subsubsection{\textbf{Attack Performance}}
To address \textbf{RQ1} and \textbf{RQ2}, we evaluate the efficacy of our attack to various target models, as follows.

\vspace{1mm}
\noindent{\bf Degree of Attack Success: }
In the context of data poisoning attacks in career prediction for online job platforms, it is important to measure how well the attacks promote or demote the target in order to measure the success rate of the attack. For this,
we use the {\em Improvement Rate} (IR) of the average target Hit Ratio (\ie, HR) in the original model as our measure. 
The improvement rate \( \text{IR} \) is defined as the increase in \( HR\) after data injection over the \( HR\) before data injection, as follows: 
\begin{equation}
\text{IR}@k = \frac{{\text{HR@k}_{\text{after}}}}{{\text{HR@k}_{\text{before}}}}
\end{equation}
% This metric is used for the promotion attack while we switch $\text{HR@10}_{\text{after}}$ with $\text{HR@10}_{\text{before}}$ for the demotion attack. 
This gives us an indication of how much we are able to manipulate the visibility of the target through data poisoning. 
% \textcolor{blue}{
For a detailed explanation and intuition of this metric, see Appendix \ref{app:IR_description}.
% } 
We vary the injection ratio in our experiments to discern its impact on the attack's success. Following previous studies \cite{yue2021black}, we set $k$=10.

\vspace{1mm}
\noindent{\bf Target Company and User Selection: }
In the company promotion and demotion attacks, we randomly sample 100 companies from ``Small'', ``Large'', and random companies on our dataset (see Section \ref{sec:Dataset} for the company definition), and measure the average $IR@10$ for the target companies. 
In the user promotion attack, we set ``Large'' companies as target companies assuming that some users want to get an interview or any recruitment opportunity for top companies competing with other job seekers, and extract users from those who never experienced ``Large'' companies (we name these as ``Specific'' users) or sample 20\% users from all users as the target users (we name this as ``Random'' users). Afterward, we see the average HR@10 for the target companies in the target users. 

\vspace{1mm}
\noindent{\bf Target Victim Models: }
To attack the career prediction models, we set the three state-of-the-art models as target victim models: NEMO \cite{li2017nemo}, AHEAD \cite{zhang2021attentive}, and NAOMI \cite{yamashita2022looking}. All three models are designed and experienced in the data from online job platforms or real-world resumes. 

\vspace{1mm}
\noindent{\bf Baseline Attack Models: }
As aforementioned, to the best of our knowledge, no existing work addresses our presented task (\ie, data poisoning attacks on career prediction). The most relevant work to ours is \cite{yue2021black}, however, their data poisoning attacks use alternating sequences (\ie, [target, non-target, target, non-target, ...]), resulting in clearly unrealistic resumes that can be easily detected by a simple rule-based system. Consequently, we compare {\ours} with existing methods that are most compatible. 
\begin{itemize}
    \item \textbf{Random:} This attack randomly generates job trajectories and inserts 1) a target company for the promotion attack and 2) a non-targeted company for the demotion attack. 
    \item \textbf{Popular:} We prepare the top 10\% frequent companies. Then, this model randomly generates job trajectories from those frequent companies and follows the same process of the random attack. 
    \item \textbf{GPT-4:} GPT-4 is the latest large language model. Due to the model's robustness and generalizability in various domains \cite{OpenAI2023GPT4TR}, we assume GPT-4  may be also useful for the HR domain. We use the zero-shot approach to obtain job trajectories. Based on the impersonation strategy, we use the following prompt to make GPT-4 generate fake trajectories. 
    % \vspace{-1pt}
    \begin{tcolorbox}[fontupper=\small]
    \textbf{Prompt}: You are a professional career advisor. I'm seeking your assistance to generate realistic career trajectories for professionals in the \{\{tech or business\}\} field. Can you provide \{\{n\}\} career paths, each containing at least five job experiences? Please ensure that all company names mentioned are real-world entities. 
    Our primary objective is to \{\{increase or reduce\}\} the likelihood of the following target companies by adding them to HR models. 
    Target Companies List: \{\{target\_company\_list\}\}
    \end{tcolorbox} 
    % \vspace{-5pt}
    Due to the output length limitation of GPT-4, we only show the injection ratio $0.1\%$ and $1\%$ for this baseline. 
    \item \textbf{DQN:} Deep Q-Network (DQN) underlies an RNN architecture tailored for sequential career trajectories. This model is trained with rewards derived from the prominence and rank of target jobs within top-k predictions \cite{mnih2013playing, zhang2020practical}. This model is used only  for the promotion attacks due to the limited nature of the loss function in the original model. 
\end{itemize}

% Prompt
% I'm seeking your assistance to generate realistic career trajectories for professionals in the tech field. Can you provide ten distinct career paths, each containing at least five job experiences? Please ensure that all company names mentioned are real-world entities.

% Our primary objective is to enhance the visibility of the following target companies by injecting them into HR models:

% Target Companies List:
% []

% For output, please just return the trajectories of company names in a Python-list compatible format.

\subsubsection{\textbf{Effect of Fake Resume Injection on Downstream Task Performance}}
Another challenge in injecting fake resumes is to make them indistinguishable from real resumes. If the overall career prediction after data poisoning changed much, the system would easily notice and alert it. Thus, to answer \textbf{RQ3}, we examine the performance shift in the career prediction before and after the data poisoning to see how much it affects the performance compared to the baseline attack models.

% \vspace{1mm}
% \noindent{\bf Network Similarity: }
% We examine the network similarity between the original resume and the fake resume, and we use graph edit distance (GED) as our metric. This measure allows us to understand the difference between the artificially created resume and the real-world resume. Specifically, we adopt the approximate algorithm \cite{abu2015exact} as GED is NP-hard to compute and thus we want to speed it up. A lower score indicates that the fake resumes align more closely with real resumes, making its detection by a recruiter or automatic detection challenging. 

% \vspace{1mm}
% \noindent{\bf Change of Downstream Task Performance: }

% We also perform fake resume detection on the dataset after injecting fake resumes to test how much harder these are to detect compared to the baseline attack models. 
% We employ a random forest classifier equipped with tf-idf features for this detection task. From each attack model, we randomly select 10\% of the fake resumes and integrate them with genuine resume datasets. After training the model, we test it using another 10\% of randomly selected fake resumes from each model and evaluate its accuracy. 

% popular/random/gpt4/mfa/ours -> 5000 from original and 1000 from each fake algo, and then tfidf vector and train -> test another 1000 from each fake algo and see acc
% small/large/random (target) * tech/biz

\subsection{Result}
\begin{table*}[t]
\centering
\caption{\textit{Company Promotion Attack} - Improvement Rate@10 of {\ours} vs baselines. The adjacent step in the reality module is three. In the promotion attack, \ul{a higher score is better}. The best and second-best results are in bold and underlined, respectively.}
\label{tab:CompanyPromotion}
\vspace{-3pt}
\begin{tabular}{P{2.15cm}|c||P{1cm}P{1cm}P{1cm}P{1cm}P{1.1cm}|P{1cm}P{1cm}P{1cm}P{1cm}P{1.1cm}}
\hline
\multirow{3}{*} & \multirow{3}{*} & \multicolumn{10}{c}{Dataset} \\
\cline{3-12}
Target Company & Injection & \multicolumn{5}{c|}{Tech} & \multicolumn{5}{c}{Business} \\
\cline{3-12}
& & Random & Popular & GPT-4 & DQN & \ours & Random & Popular & GPT-4 & DQN & \ours \\
\hline
\hline
\multirow{4}{*}{Small-Size} & 0.1\% & 0.83 & 1.10 & 0.73 & \underline{1.22} & \textbf{1.32} & 1.00 & \underline{1.08} & \underline{1.08} & 1.06 & \textbf{1.23} \\
\cline{2-12}
& 1\% & \underline{1.59} & 1.34 & 0.73 & 1.46 & \textbf{4.98} & \underline{1.90} & 1.16 & 0.83 & 1.16 & \textbf{3.56} \\
\cline{2-12}
& 5\% & \underline{4.49} & 4.24 & - & 1.56 & \textbf{15.46} & 2.56 & \underline{3.49} & - & 1.20 & \textbf{9.57} \\
\cline{2-12}
& 10\% & \underline{7.90} & 6.20 & - & 1.56 & \textbf{23.17} & 4.56 & \underline{4.99} & - & 1.23 & \textbf{10.48} \\
\hline
\multirow{4}{*}{Large-Size} & 0.1\% & \underline{1.14} & 0.98 & 1.09 & 1.09 & \textbf{1.26} & \underline{1.07} & 1.06 & \underline{1.07} & \underline{1.07} & \textbf{1.10} \\
\cline{2-12}
& 1\% & 1.33 & 1.14 & \underline{1.53} & 1.41 & \textbf{1.67} & 1.10 & 1.06 & 1.09 & \underline{1.15} & \textbf{1.39} \\
\cline{2-12}
& 5\% & \underline{1.54} & 1.45 & - & 1.36 & \textbf{3.80} & \underline{1.39} & 1.36 & - & 1.16 & \textbf{2.81} \\
\cline{2-12}
& 10\% & 1.86 & \underline{2.04} & - & 1.41 & \textbf{3.36} & 1.63 & \underline{1.65} & - & 1.28 & \textbf{2.88} \\
\hline
\multirow{4}{*}{Random-Size} & 0.1\% & 1.00 & 1.06 & \underline{1.09} & 1.04 & \textbf{1.13} & \underline{1.04} & 0.91 & 0.86 & \underline{1.04} & \textbf{1.16} \\
\cline{2-12}
& 1\% & 1.40 & 1.36 & \underline{2.27} & 1.13 & \textbf{2.49} & 1.41 & 1.23 & \underline{1.60} & 1.36 & \textbf{2.27} \\
\cline{2-12}
& 5\% & \underline{2.40} & 1.95 & - & 1.31 & \textbf{6.36} & \underline{2.89} & 2.35 & - & 1.38 & \textbf{7.90} \\
\cline{2-12}
& 10\% & 3.00 & \underline{3.40} & - & 1.31 & \textbf{9.45} & 4.37 & \underline{4.44} & - & 1.41 & \textbf{8.27} \\
\hline
\end{tabular}
\end{table*}

\begin{table*}[th]
\centering
\caption{\textit{Company Demotion Attack} - Improvement Rate@10 of {\ours} vs baselines. The adjacent step in the reality module is three. In the demotion attack, \ul{a lower score is better}. The best and second-best results are in bold and underlined, respectively.}
\label{tab:CompanyDemotion}
\vspace{-3pt}
\begin{tabular}{P{2.15cm}|c||P{1cm}P{1cm}P{1cm}P{1cm}P{1.1cm}|P{1cm}P{1cm}P{1cm}P{1cm}P{1.1cm}}
\hline
\multirow{3}{*} & \multirow{3}{*} & \multicolumn{10}{c}{Dataset} \\
\cline{3-12}
Target Company & Injection & \multicolumn{5}{c|}{Tech} & \multicolumn{5}{c}{Business} \\
\cline{3-12}
& & Random & Popular & GPT-4 & DQN & \ours & Random & Popular & GPT-4 & DQN & \ours \\
\hline
\hline
\multirow{4}{*}{Small-Size} & 0.1\% & 1.07 & \underline{0.83} & 1.00 & \scriptsize N/A & \textbf{0.73} & 1.00 & \underline{0.90} & 1.16 & \scriptsize N/A & \textbf{0.73} \\
\cline{2-12}
& 1\% & \underline{0.83} & 1.07 & 1.00 & \scriptsize N/A & \textbf{0.61} & 0.90 & 1.00 & \underline{0.83} & \scriptsize N/A & \textbf{0.75} \\
\cline{2-12}
& 5\% & 0.98 & \underline{0.83} & - & \scriptsize N/A & \textbf{0.73} & \textbf{0.73} & 0.83 & - & \scriptsize N/A & \underline{0.75} \\
\cline{2-12}
& 10\% & \underline{0.83} & \underline{0.83} & - & \scriptsize N/A & \textbf{0.61} & 1.06 & \underline{0.83} & - & \scriptsize N/A & \textbf{0.75} \\
\hline
\multirow{4}{*}{Large-Size} & 0.1\% & 1.14 & 1.19 & \textbf{0.91} & \scriptsize N/A & \underline{1.02} & \textbf{0.98} & 1.18 & 1.03 & \scriptsize N/A & \underline{0.99} \\
\cline{2-12}
& 1\% & 1.08 & \underline{1.04} & 1.06 & \scriptsize N/A & \textbf{0.98} & 1.07 & \underline{1.01} & 1.09 & \scriptsize N/A & \textbf{0.95} \\
\cline{2-12}
& 5\% & 1.08 & \underline{1.01} & - & \scriptsize N/A & \textbf{0.98} & \underline{0.95} & 1.10 & - & \scriptsize N/A & \textbf{0.94} \\
\cline{2-12}
& 10\% & 1.01 & \underline{0.98} & - & \scriptsize N/A & \textbf{0.93} & 1.01 & \underline{0.98} & - & \scriptsize N/A & \textbf{0.94} \\
\hline
\multirow{4}{*}{Random-Size} & 0.1\% & \underline{0.95} & 1.13 & 1.09 & \scriptsize N/A & \textbf{0.86} & 1.28 & \underline{0.91} & \textbf{0.86} & \scriptsize N/A & \textbf{0.86} \\
\cline{2-12}
& 1\% & \underline{0.95} & 1.04 & 1.09 & \scriptsize N/A & \textbf{0.86} & \underline{0.91} & \underline{0.91} & 1.48 & \scriptsize N/A & \textbf{0.80} \\
\cline{2-12}
& 5\% & 0.95 & \underline{0.85} & - & \scriptsize N/A & \textbf{0.82} & \textbf{0.79} & \textbf{0.79} & - & \scriptsize N/A & \underline{0.80} \\
\cline{2-12}
& 10\% & \textbf{0.85} & 0.95 & - & \scriptsize N/A & \underline{0.86} & \underline{0.79} & 0.86 & - & \scriptsize N/A & \textbf{0.68} \\
\hline
\end{tabular}
\end{table*}

\begin{table*}[th]
\centering
\caption{\textit{User Promotion Attack} - Improvement Rate@10 of {\ours} vs baselines. The adjacent step in the reality module is three, and the target company is ``Large''. As to the target users, ``Specific'' users are users who never experienced ``Large'' companies, while ``Random'' users are those randomly sampled 20\% of all users. In the promotion attack, \ul{a higher score is better}. The best and second-best results are in bold and underlined, respectively.}
\label{tab:UserPromotion}
\vspace{-3pt}
\begin{tabular}{P{2.15cm}|c||P{1cm}P{1cm}P{1cm}P{1cm}P{1.1cm}|P{1cm}P{1cm}P{1cm}P{1cm}P{1.1cm}}
\hline
\multirow{3}{*} & \multirow{3}{*} & \multicolumn{10}{c}{Dataset} \\
\cline{3-12}
Target Users & Injection & \multicolumn{5}{c|}{Tech} & \multicolumn{5}{c}{Business} \\
\cline{3-12}
& & Random & Popular & GPT-4 & DQN & \ours & Random & Popular & GPT-4 & DQN & \ours \\
\hline
\hline
\multirow{4}{*}{Specific Users} & 0.1\% & 1.00 & 1.00 & 1.03 & \textbf{1.12} & \underline{1.11} & \textbf{1.13} & 0.97 & 0.97 & \underline{1.06} & 0.98 \\
\cline{2-12}
& 1\% & 1.10 & 1.10 & \underline{1.45} & 1.22 & \textbf{1.51} & \underline{1.16} & 1.06 & 1.03 & 0.98 & \textbf{1.56} \\
\cline{2-12}
& 5\% & 1.37 & \underline{1.63} & - & 1.21 & \textbf{2.90} & \underline{1.47} & \underline{1.47} & - & 1.19 & \textbf{2.45} \\
\cline{2-12}
& 10\% & 1.93 & \underline{1.97} & - & 1.30 & \textbf{3.80} & \underline{1.72} & 1.63 & - & 1.16 & \textbf{2.48} \\
\hline
\multirow{4}{*}{Random Users} & 0.1\% & \textbf{1.24} & \underline{1.20} & 1.16 & 1.08 & 1.12 & 1.08 & 0.89 & \textbf{1.22} & 1.11 & \underline{1.17} \\
\cline{2-12}
& 1\% & \underline{1.20} & 1.08 & 1.32 & 1.08 & \textbf{2.24} & 1.03 & 1.11 & 1.09 & \underline{1.41} & \textbf{1.54} \\
\cline{2-12}
& 5\% & 1.52 & \underline{1.56} & - & 1.28 & \textbf{6.64} & \underline{1.54} & 1.32 & - & 1.39 & \textbf{2.32} \\
\cline{2-12}
& 10\% & \underline{2.40} & 1.88 & - & 1.20 & \textbf{13.08} & \underline{1.70} & 1.43 & - & 1.41 & \textbf{2.70} \\
\hline
\end{tabular}
\end{table*}

\subsubsection{\textbf{RQ1: Attack Feasibility}}
Tables \ref{tab:CompanyPromotion}, \ref{tab:CompanyDemotion}, and \ref{tab:UserPromotion} show the results of company promotion attack, company demotion attack, and user promotion attack, respectively. 
In these tables, we use NEMO \cite{li2017nemo}, LinkedIn's career prediction model, as the target victim model, and we set three steps in our reality regulation module. 

\vspace{1mm}
\noindent{\bf Overall: } 
Our results from Tables \ref{tab:CompanyPromotion}-\ref{tab:UserPromotion} clearly illustrate that, regardless of the datasets (\ie, Tech and Business), attack scenarios, target companies, injection rates, and attack methods, there are more or less vulnerabilities by data poisoning on the career prediction, with the vulnerabilities amplifying in proportion to poisoning intensity. 
Remarkably, even minimal injections, as low as 0.1\% or 1\%, can induce a significant drop in the model's expected behavior. 

\begin{figure}[t]
    \centering
    \includegraphics[width=0.73\linewidth]{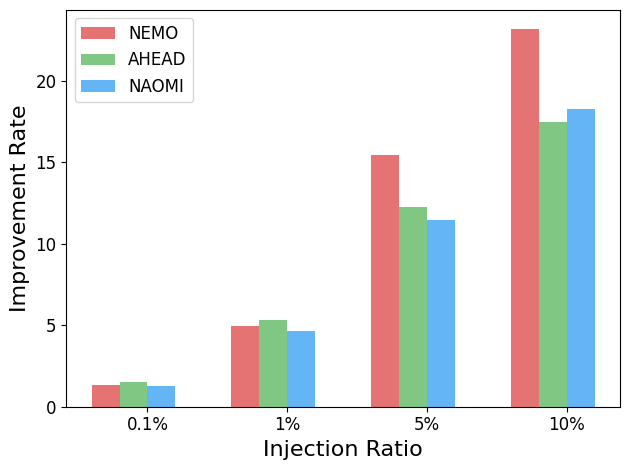}
    \vspace{-7pt}
    \caption{Victim model's improvement rate comparison in the Tech dataset with our attack method, targeting ``Small-size'' company.}
    \label{fig:victim_comparison}
    \vspace{-10pt}
\end{figure}

Figure \ref{fig:victim_comparison} shows the comparison of the improvement rate on each victim career prediction model when attacked by {\ours}. Here, we used the Tech dataset targeting ``Small'' companies. We can observe that each model can be attacked successfully, and the vulnerability increases by injection ratio. 
In the subsequent sections, we delve deeper into the discussions of each specific attack setting on NEMO \cite{li2017nemo}.

\subsubsection{\textbf{RQ2: Attack Performance Comparison}}\

\vspace{1mm}
\noindent{\bf Company Promotion Attack: } 
In this attack, a higher score indicates a better outcome, which means the attack model improves the target companies' visibility. 
The results reveal a pronounced impact when targeting ``Small'' companies with 10\% injection, and {\ours} achieved an improvement rate of 23.17 and 10.48 for the corresponding Tech and Business datasets, which 2.9 times higher than best baseline in Tech domain, and 2.1 times higher than best baseline in Business domain.  Even with a tiny injection amount of 0.1\%, {\ours} performance relatively improved 8.2\% compared to best baseline \textit{DQN} in Tech domain, and enhanced 13.9\% compared to the best baseline \textit{GPT-4} in Business domain.

When targeting ``Large'' companies, {\ours} still outperformed all the compared baselines although the impact is diminished. This reduced vulnerability can be attributed to the prevalent representation of large companies within the data, as shown in Figure \ref{fig:company_stat}. While inducing a drastic enhancement remains challenging, an attack is still attainable. 
For instance, with only 0.1\% injection, our model relatively improved 12.1\% over best baseline GPT-4 in Tech domain, and 3\% over the best baselines GPT-4 and DQN  in Business domain. The improvement is much higher with a larger injection rate. Specifically, at 5\% injection rate, our {\ours} performance is 2.8 times higher than the best baseline in Tech domain, and 2 times higher than the best baseline in Business domain.

For ``Random'' companies, we still observe the similar improvement pattern of our proposed model. At 1\% injection, our model relatively improved 9.7\% over best baseline GPT-4 in Tech domain, and 41.9\% over the best baselines DQN in Business domain.

%On the contrary, when targeting Large-Size companies, the impact is noticeably diminished. This reduced vulnerability can be attributed to the prevalent representation of large companies within the data, as shown in Figure \ref{fig:company_stat}. While inducing a drastic enhancement remains challenging, an attack is still attainable. 
%For Random-Size companies, the observed effects lie somewhere in between the aforementioned extremes.

Interestingly, while the GPT-4-induced synthetic resumes demonstrate some efficacy against Large-Size and Random-Size companies, they become counterproductive when targeting Small-Size companies. A plausible explanation for this phenomenon might be GPT-4's extensive training on job descriptions or corpora from renowned companies. Consequently, it could be under-equipped to generate convincing content for lesser-known or smaller companies.

\vspace{1mm}
\noindent{\bf Company Demotion Attack: }
In this attack, a lower score indicates a better outcome, which means the attack model reduces the target companies' visibility. 
Compared with the company promotion attack, the effects stemming from the company demotion attack are weak, but it still remains effective in manipulating prediction results. Also, we can observe that the Random attack is reasonably influential in this attack setting.

It's particularly evident that when ``Small'' companies are the target, the attack succeeds in considerably reducing the hit ratio. In contrast, attacking ``Large'' companies yields limited returns, with a 10\% data poisoning only resulting in a modest improvement rate of around 0.93 or 0.94. This resilience can be attributed to the preponderance of large companies in the dataset, rendering the model robust against attempts to degrade prediction outcomes. This observation is consistent with the earlier finding from the company promotion attack where significant improvements are elusive for ``Large'' companies. 
When targeting Random-Size companies, the resulting impact occupies the middle.

\vspace{1mm}
\noindent{\bf User Promotion Attack: } 
In this attack, a higher score indicates better. We set the target companies as ``Large'' ones, and the target users as ``Specific'' and ``Random'' users (see the detail in Section \ref{sec:protocol}). 
Promoting users via fake resume attacks is also feasible. However, minor poisoning rates, such as 0.1\%, yield minimal observable changes, while an injection of 1\% or more can notably enhance the $HR@10$ by over 1.5 times. 

It's important to note that ``Specific'' users are characterized by their lack of experience with ``Large'' companies. In contrast to ``Random'' users, the improvement rate for these ``Specific'' users is diminished. This trend can be tied back to our earlier discussions on the inherent robustness of ``Large'' companies. Conversely, examining the Tech data for ``Random'' users reveals a significant boost in the hit ratio after data poisoning. This suggests that predictions related to affiliations with giant tech companies might be heavily influenced by prior experiences with ``Large'' companies, implying {\ours} may amplify users with experience with other ``Large'' companies to be recommended to more specific large companies.

\subsubsection{\textbf{RQ3: Effect of Fake Resume Injection}}\
This section evaluates the effect of injecting fake resumes. Table \ref{tab:modelperformance} shows the overall performance change rate of career prediction before and after fake resume attacks. 
To delve deeper into the implications of fake resume injections, we conducted a series of experiments on our pre-trained career prediction model (\ie, surrogate model). Specifically, we injected fake resumes generated for company promotion attack with a 1\% injection ratio into the model and proceeded with an additional training of 20 epochs. The primary objective was to gauge the relative improvement in performance from the original metrics post-injection. 
For a holistic understanding, we also implemented a comparative baseline where no additional data was introduced but the model underwent the same additional training epochs. This scenario is denoted as ``None'' in the Table \ref{tab:modelperformance}. 

While the Random and Popular attacks achieved improvements in the hit ratio during the company promotion attack, they exemplified a significant change rate in the career prediction model's accuracy, often deviating substantially from the standard performance. The performance shifts induced by GPT-4 and DQN were not consistent and varied based on the targeted companies. On the other hand, {\ours} exhibited behaviors closely aligned with the original dataset. Notably, its improvement rate was contained within one standard deviation from the original (\ie, ``None'') improvement rate. 
This consistency in {\ours}'s performance indicates the effectiveness of our reality regulation module, suggesting that it generates resumes that are not just synthetic but also highly realistic, closely mimicking genuine career trajectories.

\begin{table}[t]
 \centering
 \caption{Relative change in career prediction accuracy after fake resume injection for the company promotion attack with a 1\% injection ratio. }
 \label{tab:modelperformance}
 \vspace{-5pt}
 \resizebox{\columnwidth}{!}{
 \begin{tabular}{c||ccc|ccc}
  \hline
  & \multicolumn{3}{c|}{Tech} & \multicolumn{3}{c}{Business} \\
  \hline
  Attack & Small & Large & Rand & Small & Large & Rand \\
  \hline
  \hline
  Random & +1.33\% & +1.47\% & +1.61\% & +1.15\% & +1.08\% & +0.95\% \\
  \hline
  Popular & +1.12\% & +1.82\% & +1.75\% & +1.01\% & +1.08\% & +0.68\% \\
  \hline
  GPT-4 & +1.33\% & +1.05\% & +1.26\% & +0.95\% & +1.01\% & +0.95\% \\
  \hline
  DQN & +1.33\% & +1.47\% & +1.05\% & +1.08\% & +0.81\% & +0.95\% \\
  \hline
  {\ours} & +1.12\% & +0.98\% & +1.05\% & +1.28\% & +1.15\% & +0.88\% \\
  \hline
  \hline
  None & +1.19\% & +1.19\% & +1.19\% & +1.01\% & +1.01\% & +1.01\% \\
  \hline
 \end{tabular}
 }
 \vspace{-10pt}
\end{table}

\vspace{-2.5pt}
\section{Limitations and Future Work} 
This study focused primarily on career positions within the realms of tech and business. It is also crucial to extend our exploration into other sectors and assess performance on datasets that encompass a mix of multiple or cross-domain genres. 
Nonetheless, our research successfully underscores the vulnerabilities introduced by data poisoning in online job platforms. 
While the focus of the current investigation was career prediction, it raises concerns about potential susceptibilities in other HR downstream tasks. In the future, it would be intriguing to scrutinize how these vulnerabilities manifest across a broader spectrum of HR applications and tasks.

\vspace{-2.5pt}
\section{Conclusion}
In this paper, we highlighted vulnerabilities in career prediction through fake resume attacks. By exploiting the flexible format of resumes and the nature of online job platforms, we demonstrated the possibility of three potential attacks: (1) company promotion attack, (2) company demotion attack, and (3) user promotion attack. We proposed a fake resume generation system that manipulates predictions through data poisoning, and showed the performance in the real-world resume datasets. 
%This exposes the risk of online job platforms being compromised by ill-intentioned users. 
This underscores the vulnerability of online job platforms to potential compromise by malicious actors.

% Acknowledgement
\vspace{-2.5pt}
\begin{acks}
This work was in part supported by NSF awards \#1934782
and \#2131144, and PSU CSRAI seed grant 2021.
\end{acks}

\bibliographystyle{ACM-Reference-Format}
\balance
\bibliography{WWW24}


\begin{thebibliography}{46}

%%% the Web address.

\ifx \showCODEN    \undefined \def \showCODEN     #1{\unskip}     \fi
\ifx \showDOI      \undefined \def \showDOI       #1{#1}\fi
\ifx \showISBNx    \undefined \def \showISBNx     #1{\unskip}     \fi
\ifx \showISBNxiii \undefined \def \showISBNxiii  #1{\unskip}     \fi
\ifx \showISSN     \undefined \def \showISSN      #1{\unskip}     \fi
\ifx \showLCCN     \undefined \def \showLCCN      #1{\unskip}     \fi
\ifx \shownote     \undefined \def \shownote      #1{#1}          \fi
\ifx \showarticletitle \undefined \def \showarticletitle #1{#1}   \fi
\ifx \showURL      \undefined \def \showURL       {\relax}        \fi
% invisible to TeX
\providecommand\bibfield[2]{#2}
\providecommand\bibinfo[2]{#2}
\providecommand\natexlab[1]{#1}
\providecommand\showeprint[2][]{arXiv:#2}

\bibitem[Ahmed and Kashmoola(2021)]%
        {ahmed2021threats}
\bibfield{author}{\bibinfo{person}{Ibrahim~M Ahmed} {and}
  \bibinfo{person}{Manar~Younis Kashmoola}.} \bibinfo{year}{2021}\natexlab{}.
\newblock \showarticletitle{Threats on machine learning technique by data
  poisoning attack: A survey}. In \bibinfo{booktitle}{\emph{Advances in Cyber
  Security: Third International Conference, ACeS 2021, Penang, Malaysia, August
  24--25, 2021, Revised Selected Papers 3}}. Springer,
  \bibinfo{pages}{586--600}.
\newblock


\bibitem[Chakraborty et~al\mbox{.}(2018)]%
        {chakraborty2018adversarial}
\bibfield{author}{\bibinfo{person}{Anirban Chakraborty},
  \bibinfo{person}{Manaar Alam}, \bibinfo{person}{Vishal Dey},
  \bibinfo{person}{Anupam Chattopadhyay}, {and} \bibinfo{person}{Debdeep
  Mukhopadhyay}.} \bibinfo{year}{2018}\natexlab{}.
\newblock \showarticletitle{Adversarial attacks and defences: A survey}.
\newblock \bibinfo{journal}{\emph{arXiv preprint arXiv:1810.00069}}
  (\bibinfo{year}{2018}).
\newblock


\bibitem[Chen et~al\mbox{.}(2017)]%
        {chen2017targeted}
\bibfield{author}{\bibinfo{person}{Xinyun Chen}, \bibinfo{person}{Chang Liu},
  \bibinfo{person}{Bo Li}, \bibinfo{person}{Kimberly Lu}, {and}
  \bibinfo{person}{Dawn Song}.} \bibinfo{year}{2017}\natexlab{}.
\newblock \showarticletitle{Targeted backdoor attacks on deep learning systems
  using data poisoning}.
\newblock \bibinfo{journal}{\emph{arXiv preprint arXiv:1712.05526}}
  (\bibinfo{year}{2017}).
\newblock


\bibitem[Dai et~al\mbox{.}(2018)]%
        {dai2018adversarial}
\bibfield{author}{\bibinfo{person}{Hanjun Dai}, \bibinfo{person}{Hui Li},
  \bibinfo{person}{Tian Tian}, \bibinfo{person}{Xin Huang},
  \bibinfo{person}{Lin Wang}, \bibinfo{person}{Jun Zhu}, {and}
  \bibinfo{person}{Le Song}.} \bibinfo{year}{2018}\natexlab{}.
\newblock \showarticletitle{Adversarial attack on graph structured data}. In
  \bibinfo{booktitle}{\emph{International conference on machine learning
  (ICML)}}. PMLR, \bibinfo{pages}{1115--1124}.
\newblock


\bibitem[Dai et~al\mbox{.}(2020)]%
        {dai2020enterprise}
\bibfield{author}{\bibinfo{person}{Le Dai}, \bibinfo{person}{Yu Yin},
  \bibinfo{person}{Chuan Qin}, \bibinfo{person}{Tong Xu},
  \bibinfo{person}{Xiangnan He}, \bibinfo{person}{Enhong Chen}, {and}
  \bibinfo{person}{Hui Xiong}.} \bibinfo{year}{2020}\natexlab{}.
\newblock \showarticletitle{Enterprise Cooperation and Competition Analysis
  with a Sign-Oriented Preference Network}. In
  \bibinfo{booktitle}{\emph{Proceedings of the 26th ACM SIGKDD International
  Conference on Knowledge Discovery \& Data Mining (KDD)}}.
  \bibinfo{pages}{774--782}.
\newblock


\bibitem[Dave et~al\mbox{.}(2018)]%
        {dave2018combined}
\bibfield{author}{\bibinfo{person}{Vachik~S Dave}, \bibinfo{person}{Baichuan
  Zhang}, \bibinfo{person}{Mohammad Al~Hasan}, \bibinfo{person}{Khalifeh
  AlJadda}, {and} \bibinfo{person}{Mohammed Korayem}.}
  \bibinfo{year}{2018}\natexlab{}.
\newblock \showarticletitle{A combined representation learning approach for
  better job and skill recommendation}. In
  \bibinfo{booktitle}{\emph{Proceedings of the 27th ACM International
  Conference on Information and Knowledge Management (CIKM)}}.
  \bibinfo{pages}{1997--2005}.
\newblock


\bibitem[Davis et~al\mbox{.}(2020)]%
        {davis2020networking}
\bibfield{author}{\bibinfo{person}{Joanna Davis}, \bibinfo{person}{Hans-Georg
  Wolff}, \bibinfo{person}{Monica~L Forret}, {and} \bibinfo{person}{Sherry~E
  Sullivan}.} \bibinfo{year}{2020}\natexlab{}.
\newblock \showarticletitle{Networking via LinkedIn: An examination of usage
  and career benefits}.
\newblock \bibinfo{journal}{\emph{Journal of Vocational Behavior}}
  \bibinfo{volume}{118} (\bibinfo{year}{2020}), \bibinfo{pages}{103396}.
\newblock


\bibitem[De~Smedt et~al\mbox{.}(2015)]%
        {de2015esco}
\bibfield{author}{\bibinfo{person}{Johan De~Smedt}, \bibinfo{person}{Martin le
  Vrang}, {and} \bibinfo{person}{Agis Papantoniou}.}
  \bibinfo{year}{2015}\natexlab{}.
\newblock \showarticletitle{ESCO: Towards a Semantic Web for the European Labor
  Market.}. In \bibinfo{booktitle}{\emph{LDOW@ WWW}}.
\newblock


\bibitem[Dillahunt et~al\mbox{.}(2021)]%
        {dillahunt2021examining}
\bibfield{author}{\bibinfo{person}{Tawanna~R Dillahunt}, \bibinfo{person}{Aarti
  Israni}, \bibinfo{person}{Alex~Jiahong Lu}, \bibinfo{person}{Mingzhi Cai},
  {and} \bibinfo{person}{Joey Chiao-Yin Hsiao}.}
  \bibinfo{year}{2021}\natexlab{}.
\newblock \showarticletitle{Examining the use of online platforms for
  employment: A survey of US job seekers}. In
  \bibinfo{booktitle}{\emph{Proceedings of the 2021 CHI conference on human
  factors in computing Systems}}. \bibinfo{pages}{1--23}.
\newblock


\bibitem[Fan et~al\mbox{.}(2022)]%
        {fan2022survey}
\bibfield{author}{\bibinfo{person}{Jiaxin Fan}, \bibinfo{person}{Qi Yan},
  \bibinfo{person}{Mohan Li}, \bibinfo{person}{Guanqun Qu}, {and}
  \bibinfo{person}{Yang Xiao}.} \bibinfo{year}{2022}\natexlab{}.
\newblock \showarticletitle{A Survey on Data Poisoning Attacks and Defenses}.
  In \bibinfo{booktitle}{\emph{2022 7th IEEE International Conference on Data
  Science in Cyberspace (DSC)}}. IEEE, \bibinfo{pages}{48--55}.
\newblock


\bibitem[Jin et~al\mbox{.}(2021)]%
        {jin2021adversarial}
\bibfield{author}{\bibinfo{person}{Wei Jin}, \bibinfo{person}{Yaxing Li},
  \bibinfo{person}{Han Xu}, \bibinfo{person}{Yiqi Wang},
  \bibinfo{person}{Shuiwang Ji}, \bibinfo{person}{Charu Aggarwal}, {and}
  \bibinfo{person}{Jiliang Tang}.} \bibinfo{year}{2021}\natexlab{}.
\newblock \showarticletitle{Adversarial attacks and defenses on graphs}.
\newblock \bibinfo{journal}{\emph{ACM SIGKDD Explorations Newsletter}}
  \bibinfo{volume}{22}, \bibinfo{number}{2} (\bibinfo{year}{2021}),
  \bibinfo{pages}{19--34}.
\newblock


\bibitem[Kong et~al\mbox{.}(2021)]%
        {kong2021survey}
\bibfield{author}{\bibinfo{person}{Zixiao Kong}, \bibinfo{person}{Jingfeng
  Xue}, \bibinfo{person}{Yong Wang}, \bibinfo{person}{Lu Huang},
  \bibinfo{person}{Zequn Niu}, {and} \bibinfo{person}{Feng Li}.}
  \bibinfo{year}{2021}\natexlab{}.
\newblock \showarticletitle{A survey on adversarial attack in the age of
  artificial intelligence}.
\newblock \bibinfo{journal}{\emph{Wireless Communications and Mobile
  Computing}}  \bibinfo{volume}{2021} (\bibinfo{year}{2021}),
  \bibinfo{pages}{1--22}.
\newblock


\bibitem[Le et~al\mbox{.}(2020)]%
        {le2020malcom}
\bibfield{author}{\bibinfo{person}{Thai Le}, \bibinfo{person}{Suhang Wang},
  {and} \bibinfo{person}{Dongwon Lee}.} \bibinfo{year}{2020}\natexlab{}.
\newblock \showarticletitle{Malcom: Generating malicious comments to attack
  neural fake news detection models}. In \bibinfo{booktitle}{\emph{2020 IEEE
  International Conference on Data Mining (ICDM)}}. IEEE,
  \bibinfo{pages}{282--291}.
\newblock


\bibitem[le~Vrang et~al\mbox{.}(2014)]%
        {le2014esco}
\bibfield{author}{\bibinfo{person}{Martin le Vrang}, \bibinfo{person}{Agis
  Papantoniou}, \bibinfo{person}{Erika Pauwels}, \bibinfo{person}{Pieter
  Fannes}, \bibinfo{person}{Dominique Vandensteen}, {and}
  \bibinfo{person}{Johan De~Smedt}.} \bibinfo{year}{2014}\natexlab{}.
\newblock \showarticletitle{Esco: Boosting job matching in europe with semantic
  interoperability}.
\newblock \bibinfo{journal}{\emph{Computer}} \bibinfo{volume}{47},
  \bibinfo{number}{10} (\bibinfo{year}{2014}), \bibinfo{pages}{57--64}.
\newblock


\bibitem[Li et~al\mbox{.}(2017a)]%
        {li2017prospecting}
\bibfield{author}{\bibinfo{person}{Huayu Li}, \bibinfo{person}{Yong Ge},
  \bibinfo{person}{Hengshu Zhu}, \bibinfo{person}{Hui Xiong}, {and}
  \bibinfo{person}{Hongke Zhao}.} \bibinfo{year}{2017}\natexlab{a}.
\newblock \showarticletitle{Prospecting the career development of talents: A
  survival analysis perspective}. In \bibinfo{booktitle}{\emph{Proceedings of
  the 23rd ACM SIGKDD International Conference on Knowledge Discovery \& Data
  Mining (KDD)}}. \bibinfo{pages}{917--925}.
\newblock


\bibitem[Li et~al\mbox{.}(2017b)]%
        {li2017nemo}
\bibfield{author}{\bibinfo{person}{Liangyue Li}, \bibinfo{person}{How Jing},
  \bibinfo{person}{Hanghang Tong}, \bibinfo{person}{Jaewon Yang},
  \bibinfo{person}{Qi He}, {and} \bibinfo{person}{Bee-Chung Chen}.}
  \bibinfo{year}{2017}\natexlab{b}.
\newblock \showarticletitle{NEMO: Next career move prediction with contextual
  embedding}. In \bibinfo{booktitle}{\emph{Proceedings of the 26th
  International Conference on World Wide Web Companion (WWW)}}.
  \bibinfo{pages}{505--513}.
\newblock


\bibitem[Liu et~al\mbox{.}(2016)]%
        {liu2016fortune}
\bibfield{author}{\bibinfo{person}{Ye Liu}, \bibinfo{person}{Luming Zhang},
  \bibinfo{person}{Liqiang Nie}, \bibinfo{person}{Yan Yan}, {and}
  \bibinfo{person}{David Rosenblum}.} \bibinfo{year}{2016}\natexlab{}.
\newblock \showarticletitle{Fortune teller: predicting your career path}. In
  \bibinfo{booktitle}{\emph{Proceedings of the AAAI conference on artificial
  intelligence (AAAI)}}, Vol.~\bibinfo{volume}{30}.
\newblock


\bibitem[Meng et~al\mbox{.}(2019)]%
        {meng2019hierarchical}
\bibfield{author}{\bibinfo{person}{Qingxin Meng}, \bibinfo{person}{Hengshu
  Zhu}, \bibinfo{person}{Keli Xiao}, \bibinfo{person}{Le Zhang}, {and}
  \bibinfo{person}{Hui Xiong}.} \bibinfo{year}{2019}\natexlab{}.
\newblock \showarticletitle{A hierarchical career-path-aware neural network for
  job mobility prediction}. In \bibinfo{booktitle}{\emph{Proceedings of the
  25th ACM SIGKDD International Conference on Knowledge Discovery and Data
  Mining (KDD)}}. \bibinfo{pages}{14--24}.
\newblock


\bibitem[Miao et~al\mbox{.}(2018)]%
        {miao2018attack}
\bibfield{author}{\bibinfo{person}{Chenglin Miao}, \bibinfo{person}{Qi Li},
  \bibinfo{person}{Lu Su}, \bibinfo{person}{Mengdi Huai},
  \bibinfo{person}{Wenjun Jiang}, {and} \bibinfo{person}{Jing Gao}.}
  \bibinfo{year}{2018}\natexlab{}.
\newblock \showarticletitle{Attack under disguise: An intelligent data
  poisoning attack mechanism in crowdsourcing}. In
  \bibinfo{booktitle}{\emph{Proceedings of the 2018 World Wide Web Conference
  (WWW)}}. \bibinfo{pages}{13--22}.
\newblock


\bibitem[Mnih et~al\mbox{.}(2013)]%
        {mnih2013playing}
\bibfield{author}{\bibinfo{person}{Volodymyr Mnih}, \bibinfo{person}{Koray
  Kavukcuoglu}, \bibinfo{person}{David Silver}, \bibinfo{person}{Alex Graves},
  \bibinfo{person}{Ioannis Antonoglou}, \bibinfo{person}{Daan Wierstra}, {and}
  \bibinfo{person}{Martin Riedmiller}.} \bibinfo{year}{2013}\natexlab{}.
\newblock \showarticletitle{Playing atari with deep reinforcement learning}.
\newblock \bibinfo{journal}{\emph{arXiv preprint arXiv:1312.5602}}
  (\bibinfo{year}{2013}).
\newblock


\bibitem[OpenAI(2023)]%
        {OpenAI2023GPT4TR}
\bibfield{author}{\bibinfo{person}{OpenAI}.} \bibinfo{year}{2023}\natexlab{}.
\newblock \showarticletitle{GPT-4 Technical Report}.
\newblock \bibinfo{journal}{\emph{ArXiv}}  \bibinfo{volume}{abs/2303.08774}
  (\bibinfo{year}{2023}).
\newblock


\bibitem[Qin et~al\mbox{.}(2023)]%
        {qin2023comprehensive}
\bibfield{author}{\bibinfo{person}{Chuan Qin}, \bibinfo{person}{Le Zhang},
  \bibinfo{person}{Rui Zha}, \bibinfo{person}{Dazhong Shen},
  \bibinfo{person}{Qi Zhang}, \bibinfo{person}{Ying Sun}, \bibinfo{person}{Chen
  Zhu}, \bibinfo{person}{Hengshu Zhu}, {and} \bibinfo{person}{Hui Xiong}.}
  \bibinfo{year}{2023}\natexlab{}.
\newblock \showarticletitle{A Comprehensive Survey of Artificial Intelligence
  Techniques for Talent Analytics}.
\newblock \bibinfo{journal}{\emph{arXiv preprint arXiv:2307.03195}}
  (\bibinfo{year}{2023}).
\newblock


\bibitem[Qin et~al\mbox{.}(2018)]%
        {qin2018enhancing}
\bibfield{author}{\bibinfo{person}{Chuan Qin}, \bibinfo{person}{Hengshu Zhu},
  \bibinfo{person}{Tong Xu}, \bibinfo{person}{Chen Zhu}, \bibinfo{person}{Liang
  Jiang}, \bibinfo{person}{Enhong Chen}, {and} \bibinfo{person}{Hui Xiong}.}
  \bibinfo{year}{2018}\natexlab{}.
\newblock \showarticletitle{Enhancing person-job fit for talent recruitment: An
  ability-aware neural network approach}. In \bibinfo{booktitle}{\emph{The 41st
  international ACM SIGIR conference on research \& development in information
  retrieval (SIGIR)}}. \bibinfo{pages}{25--34}.
\newblock


\bibitem[Ramanath et~al\mbox{.}(2018)]%
        {ramanath2018towards}
\bibfield{author}{\bibinfo{person}{Rohan Ramanath}, \bibinfo{person}{Hakan
  Inan}, \bibinfo{person}{Gungor Polatkan}, \bibinfo{person}{Bo Hu},
  \bibinfo{person}{Qi Guo}, \bibinfo{person}{Cagri Ozcaglar},
  \bibinfo{person}{Xianren Wu}, \bibinfo{person}{Krishnaram Kenthapadi}, {and}
  \bibinfo{person}{Sahin~Cem Geyik}.} \bibinfo{year}{2018}\natexlab{}.
\newblock \showarticletitle{Towards deep and representation learning for talent
  search at linkedin}. In \bibinfo{booktitle}{\emph{Proceedings of the 27th ACM
  International Conference on Information and Knowledge Management (CIKM)}}.
  \bibinfo{pages}{2253--2261}.
\newblock


\bibitem[Ruparel et~al\mbox{.}(2020)]%
        {ruparel2020influence}
\bibfield{author}{\bibinfo{person}{Namita Ruparel}, \bibinfo{person}{Amandeep
  Dhir}, \bibinfo{person}{Anushree Tandon}, \bibinfo{person}{Puneet Kaur},
  {and} \bibinfo{person}{Jamid~Ul Islam}.} \bibinfo{year}{2020}\natexlab{}.
\newblock \showarticletitle{The influence of online professional social media
  in human resource management: A systematic literature review}.
\newblock \bibinfo{journal}{\emph{Technology in Society}}  \bibinfo{volume}{63}
  (\bibinfo{year}{2020}), \bibinfo{pages}{101335}.
\newblock


\bibitem[Schwarzschild et~al\mbox{.}(2021)]%
        {schwarzschild2021just}
\bibfield{author}{\bibinfo{person}{Avi Schwarzschild}, \bibinfo{person}{Micah
  Goldblum}, \bibinfo{person}{Arjun Gupta}, \bibinfo{person}{John~P Dickerson},
  {and} \bibinfo{person}{Tom Goldstein}.} \bibinfo{year}{2021}\natexlab{}.
\newblock \showarticletitle{Just how toxic is data poisoning? a unified
  benchmark for backdoor and data poisoning attacks}. In
  \bibinfo{booktitle}{\emph{International Conference on Machine Learning
  (ICML)}}. PMLR, \bibinfo{pages}{9389--9398}.
\newblock


\bibitem[Shi et~al\mbox{.}(2020a)]%
        {shi2020learning}
\bibfield{author}{\bibinfo{person}{Baoxu Shi}, \bibinfo{person}{Shan Li},
  \bibinfo{person}{Jaewon Yang}, \bibinfo{person}{Mustafa~Emre Kazdagli}, {and}
  \bibinfo{person}{Qi He}.} \bibinfo{year}{2020}\natexlab{a}.
\newblock \showarticletitle{Learning to Ask Screening Questions for Job
  Postings}. In \bibinfo{booktitle}{\emph{Proceedings of the 43rd International
  ACM SIGIR Conference on Research and Development in Information Retrieval
  (SIGIR)}}. \bibinfo{pages}{549--558}.
\newblock


\bibitem[Shi et~al\mbox{.}(2020b)]%
        {shi2020salience}
\bibfield{author}{\bibinfo{person}{Baoxu Shi}, \bibinfo{person}{Jaewon Yang},
  \bibinfo{person}{Feng Guo}, {and} \bibinfo{person}{Qi He}.}
  \bibinfo{year}{2020}\natexlab{b}.
\newblock \showarticletitle{Salience and Market-aware Skill Extraction for Job
  Targeting}. In \bibinfo{booktitle}{\emph{Proceedings of the 26th ACM SIGKDD
  International Conference on Knowledge Discovery \& Data Mining}}.
  \bibinfo{pages}{2871--2879}.
\newblock


\bibitem[Sun et~al\mbox{.}(2019)]%
        {sun2019impact}
\bibfield{author}{\bibinfo{person}{Ying Sun}, \bibinfo{person}{Fuzhen Zhuang},
  \bibinfo{person}{Hengshu Zhu}, \bibinfo{person}{Xin Song},
  \bibinfo{person}{Qing He}, {and} \bibinfo{person}{Hui Xiong}.}
  \bibinfo{year}{2019}\natexlab{}.
\newblock \showarticletitle{The impact of person-organization fit on talent
  management: A structure-aware convolutional neural network approach}. In
  \bibinfo{booktitle}{\emph{Proceedings of the 25th ACM SIGKDD International
  Conference on Knowledge Discovery \& Data Mining (KDD)}}.
  \bibinfo{pages}{1625--1633}.
\newblock


\bibitem[Teng et~al\mbox{.}(2019)]%
        {teng2019exploiting}
\bibfield{author}{\bibinfo{person}{Mingfei Teng}, \bibinfo{person}{Hengshu
  Zhu}, \bibinfo{person}{Chuanren Liu}, \bibinfo{person}{Chen Zhu}, {and}
  \bibinfo{person}{Hui Xiong}.} \bibinfo{year}{2019}\natexlab{}.
\newblock \showarticletitle{Exploiting the contagious effect for employee
  turnover prediction}. In \bibinfo{booktitle}{\emph{Proceedings of the AAAI
  Conference on Artificial Intelligence (AAAI)}}, Vol.~\bibinfo{volume}{33}.
  \bibinfo{pages}{1166--1173}.
\newblock


\bibitem[Tong et~al\mbox{.}(2010)]%
        {tong2010vulnerability}
\bibfield{author}{\bibinfo{person}{Hanghang Tong}, \bibinfo{person}{B~Aditya
  Prakash}, \bibinfo{person}{Charalampos Tsourakakis}, \bibinfo{person}{Tina
  Eliassi-Rad}, \bibinfo{person}{Christos Faloutsos}, {and}
  \bibinfo{person}{Duen~Horng Chau}.} \bibinfo{year}{2010}\natexlab{}.
\newblock \showarticletitle{On the vulnerability of large graphs}. In
  \bibinfo{booktitle}{\emph{2010 IEEE International Conference on Data
  Mining}}. IEEE, \bibinfo{pages}{1091--1096}.
\newblock


\bibitem[Wang et~al\mbox{.}(2021)]%
        {wang2021variable}
\bibfield{author}{\bibinfo{person}{Chao Wang}, \bibinfo{person}{Hengshu Zhu},
  \bibinfo{person}{Qiming Hao}, \bibinfo{person}{Keli Xiao}, {and}
  \bibinfo{person}{Hui Xiong}.} \bibinfo{year}{2021}\natexlab{}.
\newblock \showarticletitle{Variable interval time sequence modeling for career
  trajectory prediction: Deep collaborative perspective}. In
  \bibinfo{booktitle}{\emph{Proceedings of The ACM Web Conference (WWW)}}.
  \bibinfo{pages}{612--623}.
\newblock


\bibitem[Xu et~al\mbox{.}(2020)]%
        {xu2020adversarial}
\bibfield{author}{\bibinfo{person}{Han Xu}, \bibinfo{person}{Yao Ma},
  \bibinfo{person}{Hao-Chen Liu}, \bibinfo{person}{Debayan Deb},
  \bibinfo{person}{Hui Liu}, \bibinfo{person}{Ji-Liang Tang}, {and}
  \bibinfo{person}{Anil~K Jain}.} \bibinfo{year}{2020}\natexlab{}.
\newblock \showarticletitle{Adversarial attacks and defenses in images, graphs
  and text: A review}.
\newblock \bibinfo{journal}{\emph{International Journal of Automation and
  Computing}}  \bibinfo{volume}{17} (\bibinfo{year}{2020}),
  \bibinfo{pages}{151--178}.
\newblock


\bibitem[Xu et~al\mbox{.}(2018)]%
        {xu2018measuring}
\bibfield{author}{\bibinfo{person}{Tong Xu}, \bibinfo{person}{Hengshu Zhu},
  \bibinfo{person}{Chen Zhu}, \bibinfo{person}{Pan Li}, {and}
  \bibinfo{person}{Hui Xiong}.} \bibinfo{year}{2018}\natexlab{}.
\newblock \showarticletitle{Measuring the popularity of job skills in
  recruitment market: A multi-criteria approach}. In
  \bibinfo{booktitle}{\emph{Proceedings of the AAAI Conference on Artificial
  Intelligence (AAAI)}}, Vol.~\bibinfo{volume}{32}.
\newblock


\bibitem[Yamashita et~al\mbox{.}(2022)]%
        {yamashita2022looking}
\bibfield{author}{\bibinfo{person}{Michiharu Yamashita}, \bibinfo{person}{Yunqi
  Li}, \bibinfo{person}{Thanh Tran}, \bibinfo{person}{Yongfeng Zhang}, {and}
  \bibinfo{person}{Dongwon Lee}.} \bibinfo{year}{2022}\natexlab{}.
\newblock \showarticletitle{Looking further into the future: Career pathway
  prediction}. In \bibinfo{booktitle}{\emph{Proceedings of the International
  Workshop on Computational Jobs Marketplace}}.
\newblock


\bibitem[Yamashita et~al\mbox{.}(2023)]%
        {yamashita2023james}
\bibfield{author}{\bibinfo{person}{Michiharu Yamashita},
  \bibinfo{person}{Jia~Tracy Shen}, \bibinfo{person}{Thanh Tran},
  \bibinfo{person}{Hamoon Ekhtiari}, {and} \bibinfo{person}{Dongwon Lee}.}
  \bibinfo{year}{2023}\natexlab{}.
\newblock \showarticletitle{{JAMES: Normalizing Job Titles with Multi-Aspect
  Graph Embeddings and Reasoning}}. In \bibinfo{booktitle}{\emph{2023 IEEE
  International Conference on Data Science and Advanced Analytics (DSAA)}}.
  IEEE.
\newblock


\bibitem[Yang et~al\mbox{.}(2019)]%
        {yang2019next}
\bibfield{author}{\bibinfo{person}{Jaewon Yang}, \bibinfo{person}{Qi He},
  \bibinfo{person}{How Jing}, \bibinfo{person}{Bee-Chung Chen}, {and}
  \bibinfo{person}{Liangyue Li}.} \bibinfo{year}{2019}\natexlab{}.
\newblock \bibinfo{title}{Next career move prediction with contextual long
  short-term memory networks}.
\newblock
\newblock
\newblock
\shownote{US Patent App. 15/799,396}.


\bibitem[Yuan et~al\mbox{.}(2023)]%
        {yuan2023manipulating}
\bibfield{author}{\bibinfo{person}{Wei Yuan}, \bibinfo{person}{Quoc Viet~Hung
  Nguyen}, \bibinfo{person}{Tieke He}, \bibinfo{person}{Liang Chen}, {and}
  \bibinfo{person}{Hongzhi Yin}.} \bibinfo{year}{2023}\natexlab{}.
\newblock \showarticletitle{Manipulating Federated Recommender Systems:
  Poisoning with Synthetic Users and Its Countermeasures}.
\newblock \bibinfo{journal}{\emph{arXiv preprint arXiv:2304.03054}}
  (\bibinfo{year}{2023}).
\newblock


\bibitem[Yue et~al\mbox{.}(2021)]%
        {yue2021black}
\bibfield{author}{\bibinfo{person}{Zhenrui Yue}, \bibinfo{person}{Zhankui He},
  \bibinfo{person}{Huimin Zeng}, {and} \bibinfo{person}{Julian McAuley}.}
  \bibinfo{year}{2021}\natexlab{}.
\newblock \showarticletitle{Black-box attacks on sequential recommenders via
  data-free model extraction}. In \bibinfo{booktitle}{\emph{Proceedings of the
  15th ACM Conference on Recommender Systems}}. \bibinfo{pages}{44--54}.
\newblock


\bibitem[Zhang et~al\mbox{.}(2019a)]%
        {zhang2019job2vec}
\bibfield{author}{\bibinfo{person}{Denghui Zhang}, \bibinfo{person}{Junming
  Liu}, \bibinfo{person}{Hengshu Zhu}, \bibinfo{person}{Yanchi Liu},
  \bibinfo{person}{Lichen Wang}, \bibinfo{person}{Pengyang Wang}, {and}
  \bibinfo{person}{Hui Xiong}.} \bibinfo{year}{2019}\natexlab{a}.
\newblock \showarticletitle{Job2Vec: Job title benchmarking with collective
  multi-view representation learning}. In \bibinfo{booktitle}{\emph{Proceedings
  of the 28th ACM International Conference on Information and Knowledge
  Management (CIKM)}}. \bibinfo{pages}{2763--2771}.
\newblock


\bibitem[Zhang et~al\mbox{.}(2020a)]%
        {zhang2020practical}
\bibfield{author}{\bibinfo{person}{Hengtong Zhang}, \bibinfo{person}{Yaliang
  Li}, \bibinfo{person}{Bolin Ding}, {and} \bibinfo{person}{Jing Gao}.}
  \bibinfo{year}{2020}\natexlab{a}.
\newblock \showarticletitle{Practical data poisoning attack against next-item
  recommendation}. In \bibinfo{booktitle}{\emph{Proceedings of The Web
  Conference 2020 (WWW)}}. \bibinfo{pages}{2458--2464}.
\newblock


\bibitem[Zhang et~al\mbox{.}(2019b)]%
        {zhang2019data}
\bibfield{author}{\bibinfo{person}{Hengtong Zhang}, \bibinfo{person}{Tianhang
  Zheng}, \bibinfo{person}{Jing Gao}, \bibinfo{person}{Chenglin Miao},
  \bibinfo{person}{Lu Su}, \bibinfo{person}{Yaliang Li}, {and}
  \bibinfo{person}{Kui Ren}.} \bibinfo{year}{2019}\natexlab{b}.
\newblock \showarticletitle{Data poisoning attack against knowledge graph
  embedding}. In \bibinfo{booktitle}{\emph{Proceedings of the 28th
  International Joint Conference on Artificial Intelligence (IJCAI)}}.
  \bibinfo{pages}{4853--4859}.
\newblock


\bibitem[Zhang et~al\mbox{.}(2020b)]%
        {zhang2020large}
\bibfield{author}{\bibinfo{person}{Le Zhang}, \bibinfo{person}{Tong Xu},
  \bibinfo{person}{Hengshu Zhu}, \bibinfo{person}{Chuan Qin},
  \bibinfo{person}{Qingxin Meng}, \bibinfo{person}{Hui Xiong}, {and}
  \bibinfo{person}{Enhong Chen}.} \bibinfo{year}{2020}\natexlab{b}.
\newblock \showarticletitle{Large-scale talent flow embedding for company
  competitive analysis}. In \bibinfo{booktitle}{\emph{Proceedings of The Web
  Conference 2020 (WWW)}}. \bibinfo{pages}{2354--2364}.
\newblock


\bibitem[Zhang et~al\mbox{.}(2021)]%
        {zhang2021attentive}
\bibfield{author}{\bibinfo{person}{Le Zhang}, \bibinfo{person}{Ding Zhou},
  \bibinfo{person}{Hengshu Zhu}, \bibinfo{person}{Tong Xu},
  \bibinfo{person}{Rui Zha}, \bibinfo{person}{Enhong Chen}, {and}
  \bibinfo{person}{Hui Xiong}.} \bibinfo{year}{2021}\natexlab{}.
\newblock \showarticletitle{Attentive heterogeneous graph embedding for job
  mobility prediction}. In \bibinfo{booktitle}{\emph{Proceedings of the ACM
  SIGKDD International Conference on Knowledge Discovery and Data Mining
  (KDD)}}. \bibinfo{pages}{2192--2201}.
\newblock


\bibitem[Zhang et~al\mbox{.}(2020c)]%
        {zhang2020online}
\bibfield{author}{\bibinfo{person}{Xuezhou Zhang}, \bibinfo{person}{Xiaojin
  Zhu}, {and} \bibinfo{person}{Laurent Lessard}.}
  \bibinfo{year}{2020}\natexlab{c}.
\newblock \showarticletitle{Online data poisoning attacks}. In
  \bibinfo{booktitle}{\emph{Learning for Dynamics and Control}}. PMLR,
  \bibinfo{pages}{201--210}.
\newblock


\bibitem[Z{\"u}gner et~al\mbox{.}(2018)]%
        {zugner2018adversarial}
\bibfield{author}{\bibinfo{person}{Daniel Z{\"u}gner}, \bibinfo{person}{Amir
  Akbarnejad}, {and} \bibinfo{person}{Stephan G{\"u}nnemann}.}
  \bibinfo{year}{2018}\natexlab{}.
\newblock \showarticletitle{Adversarial attacks on neural networks for graph
  data}. In \bibinfo{booktitle}{\emph{Proceedings of the 24th ACM SIGKDD
  international conference on knowledge discovery \& data mining (KDD)}}.
  \bibinfo{pages}{2847--2856}.
\newblock


\end{thebibliography}

\clearpage
\nobalance
\appendix
\section{Ethics Statement}\label{app:ethics}
To elucidate our dataset and address potential ethical concerns, it is important to note that we did not scrape the web to collect out dataset. Instead, our dataset was donated from FutureFit AI, a job platform company, under a Memorandum of Understanding (MOU). Given the terms of this agreement, we are unable to publicly release the dataset. However, we are open to sharing it for research purposes with entities that submit legitimate requests (\eg, a signed MOU), provided there is a commitment to not attempt any de-anonymization of the data.

\section{Dataset Statistics}\label{app:data_stat}
\begin{figure}[tbh]
    \centering
    \includegraphics[width=\linewidth]{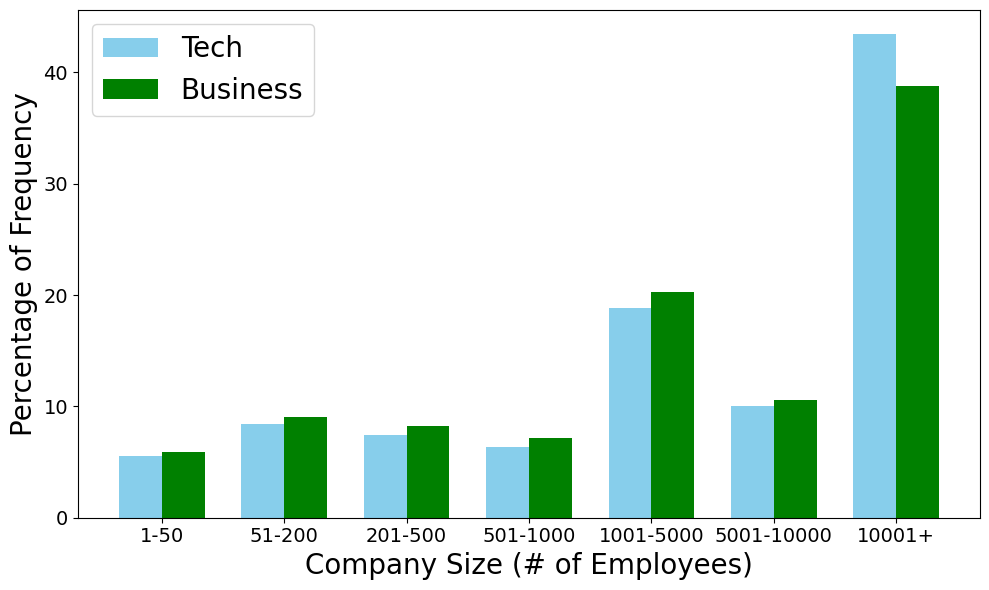}
    % \vspace{-5pt}
    \caption{Distribution of company (\# of employees).}
    \label{fig:company_stat}
    % \vspace{-5pt}
\end{figure}

In our dataset, we labeled companies as ``Small'' if they have less than 200 employees, and as ``Large'' if they have more than 10,000 employees. Figure \ref{fig:company_stat} shows the distribution of companies by size within our dataset. We show the percentage of the sum number of companies experienced by users in the Tech and Business datasets.

\section{Potential Victims in Fake Resume Attacks}\label{app:victim_description} 
To clarify our attack scenarios and provide concrete examples of the potential victims, we outline the potential attackers and victims. 

\begin{enumerate}
    \item Company Promotion Attack: 
In this scenario, smaller or emerging companies (\eg, startups), which may not be well-recognized by the system, could attack to boost their visibility, impacting the performance of job platforms as victims. 

\item Company Demotion Attack:
This type of attack could see larger companies targeting their competitors to decrease their online visibility, thereby potentially affecting the competitors' long-term financial outcomes as victims.

\item User Promotion Attack:
The attackers might be job applicants or entities lacking the requisite qualifications, aiming to increase their chances of securing interviews. The victims in this case are the recruiting companies, which risk overlooking genuinely qualified candidates due to manipulated candidate rankings.
\end{enumerate}

\begin{comment}

\vspace{1mm}
\noindent{(1) Company Promotion Attack}\\
In this scenario, smaller or emerging companies (\eg, startups), which may not be well-recognized by the system, could attack to boost their visibility, impacting the performance of job platforms as victims. 

\vspace{1mm}
\noindent{(2) Company Demotion Attack}\\
This type of attack could see larger companies targeting their competitors to decrease their online visibility, thereby potentially affecting the competitors' long-term financial outcomes as victims.

\vspace{1mm}
\noindent{(3) User Promotion Attack}\\
Here, the attackers might be job applicants or entities lacking the requisite qualifications, aiming to increase their chances of securing interviews. The victims in this case are the recruiting companies, which risk overlooking genuinely qualified candidates due to manipulated candidate rankings.

\end{comment}

\section{Further Description of Attack Feasibility}\label{app:IR_description}
\subsection{Intuition of Improvement Rate}
To provide a deeper understanding of our metric, we provide a detailed explanation and intuition of this metric. 
The improvement rate, introduced in Section \ref{sec:protocol} of our paper, refers to the extent to which the original Hit Ratio is enhanced by the attack. In another word, it can be linked as the visibility improvement of the targeted company, \ie, in case of company promotion attack (other attacking scenarios are interpreted similarly). For instance, if a small company using the best attacking baseline initially is visible to 10k users, ``an improvement rate of 1.32'' or ``32\% improvement'' means that the company would now be reached out to 3,200 more users on an online job platform via its career recommendation service. This increase, although sometimes seemingly modest, can be significant in the context of career platforms, where visibility is crucial.

\subsection{Extremely Lower Injection Rates}
We understand the practical feasibility of our proposed attacks, especially considering the large number of fake profiles required. First, we would like to emphasize that our study primarily aimed to highlight potential vulnerabilities in online job platforms. Our intent was to provoke thought and initiate a conversation about potential security enhancements in job platforms.

While our study used LinkedIn as a reference point, our findings have significant implications for a wide range of job platforms, including career-focused sites and domain-specific job platforms. The diversity in the scale and focus of these platforms underscores the importance of our research in highlighting potential vulnerabilities. The smaller scale of some platforms could make them more susceptible to the types of attacks we have outlined.

Regarding the injection rate, our Tech and Business datasets, provided by a job platform company, each contains around 10,000 resumes. A 0.1\% injection rate translates to only creating 10 fake accounts. We vary the injection rate from 0.1\% to 10\% to present how our attacking method works in different settings, and we acknowledge that 0.1\% is more practical. 
% Applying a 0.1\% injection rate to a job platform having 1 billion accounts can translate to 1 million fake resumes, which might be not practical. However, the injection rate does not necessarily scale linearly with the scale of job platform accounts while maintaining similar attacking success. In contrast, the larger scale of the job dataset can lead to a larger scale of job-transition graph, which is more vulnerable \cite{tong2010vulnerability}.

To see extremely lower injection rates (\ie, 0.01\% and 0.05\%), we conduct the same experiment on a combined Tech and Business dataset, which contains 20k resumes for the company promotion attack. Table \ref{tab:additional_ex} shows the result of this additional experiment. Here, we see that even a 10 times smaller injection rate of 0.01\%, which translates to creating merely 2 fake resumes, was effective and even better than each single Tech/Business dataset. In this experiment, our company promotion attacking method has an improvement rate of 1.14 with 0.01\% injection, and 1.58 with 0.1\% injection rate. 
% We also observe similar experimental patterns for company demotion attack and user promotion attack.

% While the attack on an extremely huge platform can be more explored, our study's focus remains on exposing vulnerabilities in the online job market ecosystem, aiming to catalyze the development of more robust defenses against such data attacks.
The reason we did not initially explore a 0.01\% injection rate was due to the constraints of our dataset size. However, with a larger dataset, it becomes possible to observe the effects of even a very small injection rate. This is because the larger dataset offers a more comprehensive representation of real-world scenarios, allowing the subtle influences of well-crafted fake resumes to become more apparent and impactful. 
To further substantiate this aspect, we refer to the study \cite{tong2010vulnerability}, which highlights that well-connected graphs with many loops and paths are more vulnerable (\ie, it is easier for a virus to propagate across the graph = the graph is less robust to the virus attack), paralleling our observation in the larger dataset where the node degree of small-size companies is higher (5.20) compared to the original Tech and Business datasets (4.72 and 4.60, respectively). This node degree change aligns with the aforementioned study's findings \cite{tong2010vulnerability}.

\begin{table}[tb]
\centering
\caption{Additional experiment result of company promotion attack. We combined both our Tech and Business datasets and conducted the same experiment.}
\label{tab:additional_ex}
\begin{tabular}{c|c}
\hline
Injection & Improvement Rate \\
\hline
0.01\% & 1.14 \\
0.05\% & 1.30 \\
0.1\% & 1.58 \\
1\% & 5.81 \\
5\% & 16.79 \\
10\% & 19.84 \\
\hline
\end{tabular}
\end{table}

\subsection{Attack Practicality}
In this section, we discuss the practical feasibility of our proposed attacks. The effectiveness depends on the attackers' objectives and available resources, which can vary widely. For instance, small companies or startups may have different objectives and constraints compared to larger companies. 

It is crucial to recognize that in addition to Linkedin, numerous small to medium scale job platforms exist worldwide, with user bases ranging from a few thousand to several millions. Moreover, these platforms often incorporate variables such as geographic locations, sectors, and job categories, which can significantly narrow down the pool of relevant accounts when these factors are taken into account.
For example, on a platform with 20,000 users, injecting just a few well-crafted fake resumes could lead to an increase in visibility. For lesser-known small companies or startups, this enhanced visibility can be considered impactful. Referencing Table \ref{tab:additional_ex}, an improvement rate of 1.14 with 2 fake resumes, or even 1.58 with 20 fake resumes in a job platform of 20,000 users, increasing 14\% or 58\% more visibility to the users, can indeed be considered effective, and this scale is not only more feasible but also practically achievable, highlighting the potential risk even at smaller scales.

% In addition, there are factors like countries, regions, job categories, etc., that are accounted for in job platform systems, which in turn scale down the number of accounts when considering these factors. 
% For instance, Bizreach – a leading job platform in some countries, has only 1 million of registered accounts and a 0.01\% injection rate would create only 100 fake profiles. 
% Considering that there are a lot of job platforms ranging from a few thousand to millions of users in the world, this scale is not only more feasible but also practically achievable, highlighting the potential risk even at smaller scales.

\section{Generalizability Discussion}\label{app:generalizability}
In this section, we discuss the generalizability of our findings. Our datasets encompass both Tech and Business sectors, demonstrating the susceptibility of these areas to fake resume attacks. 
Our study's relevance is further bolstered by the existence of domain-specific job platforms, such as those focusing exclusively on tech recruitment. The structure of user job trajectories or resumes used in our study is widely adopted by a broad range of job platforms, including career-focused sites and domain-specific job platforms. This suggests that our findings have a broad applicability.

In our discussion, we consider the potential variations in vulnerability levels across different platforms and sectors, identifying several key factors that may influence these vulnerability variations. 

\noindent {\bf(1) Platform Size and User Base:} Larger platforms might implement more sophisticated security frameworks, potentially diminishing their susceptibility, whereas smaller platforms might be inherently more vulnerable due to limited protective measures.

\noindent {\bf(2) Sector-Specific Dynamics:} Different sectors, such as legal and healthcare domains, may exhibit unique vulnerabilities owing to their distinct recruitment practices and hiring patterns. 

To encapsulate, our study sheds light on the pervasive risk of fake resume submissions across multiple sectors and job platforms, emphasizing the need for increased vigilance and improved security measures. These findings encourage both industry-specific and general job platforms to reassess and fortify their defenses against such fraudulent activities, ensuring a safer and more trustworthy platform. Our comprehensive analysis and discussion are intended to contribute to a deeper understanding of these threats, provoke thought and initiate a conversation about potential security enhancements in job platforms.

\section{User Demotion Attack}\label{app:userdemotion}
\begin{table*}[!tbh]
\centering
\caption{\textit{User Demotion Attack} - Improvement Rate@10 of {\ours} vs baselines. The adjacent step in the reality module is three, and the target company is ``Large''. As to the target users, ``Specific'' users are users who experienced ``Large'' companies at least once, while ``Random'' users are those randomly sampled 20\% of all users. In the demotion attack, \ul{a lower score is better}. The best and second-best results are in bold and underlined, respectively.}
\label{tab:UserDemotion}
\vspace{-3pt}
\begin{tabular}{P{2.15cm}|c||P{1cm}P{1cm}P{1cm}P{1cm}P{1.1cm}|P{1cm}P{1cm}P{1cm}P{1cm}P{1.1cm}}
\hline
\multirow{3}{*}{Target Users} & \multirow{3}{*}{Injection} & \multicolumn{10}{c}{Dataset} \\
\cline{3-12}
& & \multicolumn{5}{c|}{Tech} & \multicolumn{5}{c}{Business} \\
\cline{3-12}
& & Random & Popular & GPT-4 & DQN & \ours & Random & Popular & GPT-4 & DQN & \ours \\
\hline
\hline
\multirow{4}{*}{Specific Users} & 0.1\% & 1.12 & 1.33 & \underline{1.05} & \scriptsize N/A & \textbf{0.98} & 1.07 & \textbf{0.97} & \underline{1.00} & \scriptsize N/A & 1.05 \\
\cline{2-12}
& 1\% & \underline{1.05} & \underline{1.05} & 1.07 & \scriptsize N/A & \textbf{0.91} & 1.15 & 1.15 & \underline{1.05} & \scriptsize N/A & \textbf{1.00} \\
\cline{2-12}
& 5\% & \underline{0.98} & 1.05 & - & \scriptsize N/A & \textbf{0.93} & 1.07 & \underline{1.04} & - & \scriptsize N/A & \textbf{1.02} \\
\cline{2-12}
& 10\% & 1.00 & \underline{0.93} & - & \scriptsize N/A & \textbf{0.84} & \underline{0.93} & \textbf{0.90} & - & \scriptsize N/A & 0.96 \\
\hline
\multirow{4}{*}{Random Users} & 0.1\% & \textbf{0.77} & 1.09 & 1.06 & \scriptsize N/A & \underline{0.97} & 1.06 & \underline{1.00} & \textbf{0.94} & \scriptsize N/A & 1.06 \\
\cline{2-12}
& 1\% & \underline{0.97} & \textbf{0.91} & 1.03 & \scriptsize N/A & \underline{0.97} & 1.03 & \underline{0.97} & \textbf{0.94} & \scriptsize N/A & 1.00 \\
\cline{2-12}
& 5\% & \underline{0.83} & \textbf{0.80} & - & \scriptsize N/A & 0.94 & \textbf{0.97} & \underline{1.03} & - & \scriptsize N/A & 1.06 \\
\cline{2-12}
& 10\% & \textbf{0.71} & 1.00 & - & \scriptsize N/A & \underline{0.91} & 0.97 & \underline{0.92} & - & \scriptsize N/A & \textbf{0.91} \\
\hline
\end{tabular}
\end{table*}

While our architecture is capable of supporting user demotion attacks as we just need to change the loss function, we did not focus on the user-demotion attack in our main discussion as we thought such an attack was much less likely to happen. That is, while it can happen in theory (\ie, User A attacks User B so that User B becomes less likely to be recruited), we consider that promoting companies/users and demoting companies are more likely scenarios. Demoting certain users does not directly benefit attackers in terms of visibility or competitive advantage. Our emphasis was on attacks that have clear motives and tangible benefits for attackers, such as company promotion and demotion, and user promotion. 

However, in this section, we show the user demotion attack case for the sake of completeness. The objective functions of this user demotion attack can be formulated:
\begin{equation}
L_{\text{user-demotion}} = \frac{1}{U} \sum_{i \in U} \sum_{j \in T} P_{ij}
\end{equation}
In this attack, the aim is to decrease the likelihood of specific users (or resumes) being associated with target companies, optimizing over a select group of users. First, we set ``Large'' companies as target companies. Then, for the target users, we extract users from those who experienced ``Large'' companies at least once (we name these as ``Specific'' users) or sample 20\% users from all users as the target users (we name this as ``Random'' users). Afterward, we see the average HR@10 for the target companies in the target users. 

Table \ref{tab:UserDemotion} shows the results of the user demotion attack, where we use NEMO \cite{li2017nemo} as the target victim model and set three steps in our reality regulation module. In this attack, a lower score indicates better. Compared to the company demotion attack in Table \ref{tab:CompanyDemotion}, the impact of the attack is relatively lower, but it shows that it is also possible to demote users through fake resume attacks.

\end{document}